\documentclass[]{tJOT2e}
\usepackage{pifont}
\usepackage{graphicx}

\newcommand{\eqnlab}[1]{\label{eq:#1}}
\newcommand{\figlab}[1]{\label{fig:#1}}

\newcommand{\eqnref}[1]{(\ref{eq:#1})}
\newcommand{\figref}[1]{\ref{fig:#1}}

\newcommand{\Eqnref}[1]{Eq.~(\ref{eq:#1})}
\newcommand{\Figref}[1]{Fig.~\ref{fig:#1}}

\newcommand{\applab}[1]{\label{app:#1}}
\newcommand{\seclab}[1]{\label{sec:#1}}
\newcommand{\sseclab}[1]{\label{ssec:#1}}

\newcommand{\secref}[1]{\ref{sec:#1}}

\newcommand{\Appref}[1]{Appendix~\ref{app:#1}}
\newcommand{\Secref}[1]{Section~\ref{sec:#1}}
\newcommand{\Ssecref}[1]{Subsec.~\ref{ssec:#1}}

\newcommand\nn{\nonumber}
\newcommand{\sign}{\mathrm{sign}}
\newcommand{\ku}{\ensuremath{\mbox{Ku}}}
\newcommand{\st}{\ensuremath{\mbox{St}}}
\newcommand{\const}{\ensuremath{{\rm const.}}}
\newcommand{\ve}[1]{\ensuremath{\mbox{\boldmath$#1$}}}
\newcommand{\ma}[1]{\ensuremath{\mathbb{#1}}}
\newcommand{\e}{\ensuremath{{\rm e}}}
\newcommand{\muc}{\ensuremath{{\mu_{\rm c}}}}
\newcommand{\Rstar}{\ensuremath{{R^*}}}
\newcommand{\zstar}{\ensuremath{{z^*}}}
\newcommand{\zstarx}[1]{\ensuremath{{z_{#1}^*}}}
\newcommand{\xstar}{\ensuremath{{\Delta x^*}}}
\newcommand{\Vr}{\ensuremath{{V_{R}}}}
\newcommand{\dotVr}{\ensuremath{{{\dot V}_{R}}}}
\newcommand{\R}{\ensuremath{{R}}}
\newcommand{\V}{\ensuremath{{V}}}
\newcommand{\Z}{\ensuremath{{|\ve z|}}}
\newcommand{\Zleft}[1]{\ensuremath{{Z_{#1}^{({\rm l})}}}}
\newcommand{\Rc}{\ensuremath{{{|\Delta\ve x_{\rm c}|}}}}
\newcommand{\Vc}{\ensuremath{{{V_{\rm c}}}}}
\newcommand{\I}{\mbox{\raisebox{-0.05cm}{\large\ding{192}}}}
\newcommand{\II}{\mbox{\raisebox{-0.05cm}{\large\ding{193}}}}
\newcommand{\III}{\mbox{\raisebox{-0.05cm}{\large\ding{194}}}}

\newcommand{\fI}{\ensuremath{{{f_{1}}}}}
\newcommand{\fII}{\ensuremath{{{f_{2}}}}}
\newcommand\transpose{^{\mathrm T}}

\begin{document}
\doi{10.1080/14685248.YYYYxxxxxx}
 \issn{1468-5248}
 \jvol{00} \jnum{00} \jyear{2011}

\markboth{K. Gustavsson and B. Mehlig}{Journal of Turbulence}

\title{Relative velocities of inertial particles in turbulent aerosols}
\author{K. Gustavsson and B. Mehlig$^{\ast}$\thanks{$^\ast$Corresponding author.
Email: Bernhard.Mehlig@physics.gu.se\vspace{6pt}}\\
{\em Department of Physics, Gothenburg University, 41296 Gothenburg, Sweden}}

\maketitle

\begin{abstract}
We compute the joint distribution of relative velocities and separations
of identical inertial particles suspended in randomly mixing and turbulent flows.
Our results are obtained by matching asymptotic forms of the distribution.
The method takes into account spatial clustering of the suspended particles as well
as singularities in their motion (so-called \lq caustics'). It thus takes proper account
of the fractal properties of phase space and the distribution is characterised
in terms of the corresponding phase-space fractal dimension $D_2$.
The method clearly exhibits universal aspects of the distribution
(independent of the statistical properties of the flow): at small particle separations $\R$ and not too large radial
relative speeds $|\Vr|$, the distribution of radial relative
velocities exhibits a universal power-law form $\rho(\Vr,\R) \sim |\Vr|^{D_2-d-1}$
provided that $D_2 \le d+1$ and that the Stokes number $\st$ is large enough for caustics to form.
The range in $\Vr$ over which this power law is valid
depends on $\R$, on the Stokes number, and upon the nature of the flow.
Our results are in good agreement with results of computer simulations of the dynamics of particles
suspended in random velocity fields with finite correlation times. In the white-noise
limit the results are consistent with those of [Gustavsson and Mehlig, Phys. Rev. E84 (2011) 045304].

\end{abstract}

\newpage \section{Introduction}
\label{sec:I}

Collision velocities of particles in randomly mixing or turbulent flows (\lq turbulent aerosols')
have been studied intensively for several decades.  This is an important topic because the stability
of turbulent aerosols is determined by collisions between the suspended particles. One example
 is the problem of rain initiation in turbulent cumulus clouds. It is argued \cite{Dev12} that
small-scale turbulent stirring increases the collision rate of microscopic water droplets
causing them to coalesce more often and thus accelerating the growth of rain droplets.
This idea goes back to Smoluchowski \cite{Smo17}. Saffman and Turner \cite{Saf56} invoked this principle to estimate
the geometrical collision rate of small water droplets advected in the air flow of turbulent cumulus clouds,
assuming that the droplets are swept towards each other by essentially time-independent turbulent shears
(effects due to the unsteadiness of turbulent flows are discussed by
Andersson {\em et al.} \cite{And07} and Gustavsson {\em et al.} \cite{Gus08}). 

It is
now well known that particle inertia may have a substantial
effect on the collision rate.  Unlike advected particles, inertial particles are not constrained to follow the flow.  
Direct numerical simulations of particles in turbulent
flows \cite{Sun97,Wan00} show that the average collision speed (and thus the collision rate)
increases rapidly as the \lq Stokes number' $\st$ is varied beyond a threshold.
The Stokes number is a dimensionless measure of the particle inertia.
This sensitive dependence of the collision speed (and collision rate) upon $\st$ was
explained in~\cite{Wil06} (see also \cite{Fal02}) by
the occurrence of so-called \lq caustics\rq{}, singularities in the particle dynamics at non-zero
values of $\st$.  Caustics appear when  phase-space manifolds
describing the dependence of particle velocity upon particle position
fold over \cite{Wil03,Wil05,Cri92}. In the fold region, the velocity field at a given point
in space becomes multi-valued, allowing for large velocity differences between nearby
particles.  In the absence of such singularities,
in single-valued smooth particle-velocity fields,
the relative velocity of two particles tends to zero as
they approach each other. In the presence of caustics,
by contrast, the relative velocity of two colliding particles may be large.
There are now a number of different parameterisations of
the average geometrical collision
rate between inertial particles \cite{Fal02,Bec05,Chu05,Wil06},
and for average relative velocities \cite{Vol80,Wei93,Zai03,Meh07} 
conditional on a small separation $R$.

The geometrical collision rate neglects the fact that particles approaching each other may not collide:
viscous effects can cause the particles to avoid each other \cite{Pru97}. This effect is commonly
parameterised in terms of a \lq collision efficiency\rq{}. Relatively little
is known about the collision efficiency in turbulent flows. It
must sensitively depend on the relative speed of the particles.
When the relative speed is small the particles may spend a substantial amount of time close together. This
gives viscous forces time to affect the collision dynamics. The collision efficiency
depends strongly on the Stokes number and the turbulence intensity, its
mean value may range over several orders of magnitude \cite{Jon96}, and its instantaneous values fluctuate substantially \cite{Pin07}.
In order to understand these properties of the collision efficiency it is necessary to study the distribution of relative velocities in turbulent flows.
A theory of the average relative velocity is not sufficient.
The distribution of collision velocities is important also for another aspect of
the problem, namely the question under which circumstances colliding droplets
coalesce. The \lq coalescence efficiency\rq{} \cite{Dev12} may depend sensitively
on the actual collision velocity.

A second example where the distribution of relative velocities of inertial particles is of great significance
is the problem of planet formation. It is thought that the planets in our solar system
have formed out of microscopic dust grains suspended in the turbulent gas flow surrounding the sun.
It is assumed that micron-sized grains grow by collisional aggregation
in the first stage of this process. The standard model is reviewed by Youdin \cite{You08} and Armitage \cite{Arm07}
(an alternative model for planet formation was proposed by Wilkinson {\em et al.} \cite{Wil13}).
A problem in the standard model is that as the aggregates grow their Stokes number increases.
This in turn implies that the aggregates collide at larger impact velocities which
may cause the aggregates to fragment upon impact, hindering further growth.
Wilkinson {\em et al.} \cite{Wil08} discuss this problem in detail, see also \cite{Brau08}. How severe this barrier
to further growth is depends on how easily the aggregates fragment.
Several different models for the fragmentation process have been suggested \cite{Dom97,You04,Gut12}. The models have in common
that the time the aggregates spend close to each other is an important (and unknown) factor.
Even when colliding grains do not shatter, they may erode, or bounce off each other \cite{Zso10}.
The process is further complicated by the fact that the aggregates are not compact \cite{Wur98,Blu00,Kra04}.
In order to describe the kinetics of the aggregation (and fragmentation) process
it is necessary to know the distribution of relative velocities \cite{Win12}.
It is not sufficient to estimate the average relative speed of nearby particles \cite{Wei93}.

The joint distribution of relative velocities and spatial separations is determined by
the dynamics of the suspended particles in phase space. To compute this dynamics is a very complicated problem. It is commonly
simplified by assuming that the particles are identical, spherical and very small, and that they do not
directly interact with each other. In this case, provided that the particles
are larger than the mean free path of the fluid,
the equation of motion was derived
by Maxey and Riley \cite{Max83}. The problem is often further simplified by keeping only Stokes' force in the Maxey-Riley equation:
\begin{equation}
\label{eq:1}
\dot {\ve x} = \ve v\,,\quad\dot{\mbox{\boldmath$v$}}=\gamma(\mbox{\boldmath$u$}(\mbox{\boldmath$x$},t)-\mbox{\boldmath$v$})\,.
\end{equation}
Here $\mbox{\boldmath$x$}$ and $\mbox{\boldmath$v$}$ are particle position and velocity,
dots denote time derivatives, $\gamma$ is the rate at which the inertial
motion is damped relative to the fluid, and $\mbox{\boldmath$u$}(\mbox{\boldmath$x$},t)$ is the velocity field of the flow. For particles suspended
in a dilute gas (when the particles are smaller
than the mean free path of the gas, the Epstein limit)
a law of the same form as Eq.~(\ref{eq:1}) holds.

Even for Eq.~(\ref{eq:1}) the distribution of relative velocities of inertial particles was not known
until recently. We computed the joint distribution of relative velocities and spatial separations
in random velocity fields in the white-noise limit by means of diffusion approximations \cite{Gus11b}. We found that
the relative-velocity distribution at small spatial separations could be very broad (of power-law form)
and showed that this is a consequence of the existence of caustics and fractal clustering
in phase space at finite Stokes number $\st = (\gamma \tau)^{-1}$
(here $\tau$ is the correlation time of the flow, the Kolmogorov time in turbulent flows).

A second dimensionless parameter, the Kubo number $\ku = u_0\tau/\eta$ \cite{Dun05,Wil07}, is formed out of the typical flow speed $u_0$ and the correlation length $\eta$ (the Kolmogorov length).
The white-noise limit corresponds to the limit $\ku\rightarrow 0$ and $\st \rightarrow \infty$
so that $\ku^2 \st$ remains constant.

But the turbulent flow seen by a moving particle is not a white-noise signal, turbulent
flows have Kubo numbers of order unity.
Gustavsson {\em et al.} \cite{Gus12} described numerical results for the moments of relative velocities
at small separations in a kinematic model of turbulence with $\ku \sim 1$.
At very small separations, where caustics make a substantial contribution, these
numerical results are well described by the white-noise results of Ref.~\cite{Gus11b}, see also \cite{Cen11}.
It thus seems that the white-noise approximation,
based on a perturbative solution of a Fokker-Planck equation, describes important
properties of relative velocities in turbulent aerosols.
But there is to date no theory for the distribution of relative velocities for the physically most relevant case
of $\ku\sim 1$.
\begin{figure}
\includegraphics[width=12cm,clip]{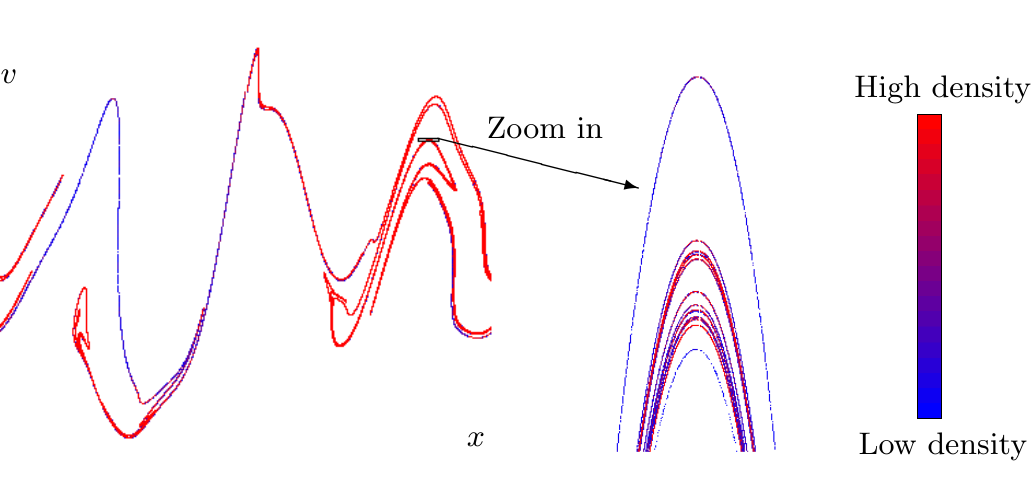}
\caption{\label{fig:attractor}
Snapshot of particle positions and velocities in one spatial dimension according to the model described in \Secref{model}.
The parameters are: $\ku=0.1$ and $\st=100$ (that is $\epsilon^2=3$ according to Eq.~\eqnref{epsilon_def}), and the system size is $L=10\eta$.
The corresponding fractal dimensions are: Kaplan-Yorke 
dimension $D_1\approx 1.07$, phase-space 
correlation dimension $D_2\approx 0.24$.
The particle number density is colour coded.}
\end{figure}

In this paper we describe a general principle determining the joint distribution
of relative velocities and spatial separations in turbulent aerosols.
In its most general form it does not rely upon the white-noise approximation.
The method is based on matching asymptotic forms of the distribution function
in phase space.

As mentioned above there are two distinct ways in which particles move relative to each other.
This gives rise to two contributions to the distribution.  First,
the formation of caustics allows particles to rapidly approach  each other on different branches
of the phase-space attractor (illustrated in Fig.~\ref{fig:attractor} in one spatial dimension).
The corresponding phase-space distribution is of power-law form (reflecting fractal clustering
in phase space). Second, if two particles approach on the same branch,
then the pair diffuses in a correlated manner and the particles stay close to each other
for a long time (determined by the inverse maximal Lyapunov exponent), at small relative velocities. It turns out that
the corresponding relative velocity distribution is approximately constant, that is
independent of the relative velocity at small separations.
Our result for the distribution of relative
velocities (\Eqnref{rho_asym_generaldim} in \Secref{3}) is obtained by glueing theses two pieces together.
This method takes into account fractal clustering in phase space and
caustic formation.

Our method makes it possible to identify which properties of the distribution function
are universal, and which properties are system specific (depend on the details of
the flow statistics). The power-law form of the distribution of relative velocities
at small separations is a universal feature. It pertains to systems that exhibit fractal clustering,
at sufficiently high Stokes numbers so that caustics are abundant. 

Our result for the distribution of relative velocities, \Eqnref{rho_asym_generaldim}, 
contains two system-specific parameters: the phase-space correlation dimension $D_2$ and a matching parameter $\zstar$.
The matching scale $\zstar$ distinguishes the two types of relative motion (due to pair diffusion and caustics) discussed above.
For small Stokes numbers caustics are rare and the dynamics is dominated by system-specific
pair diffusion. In this limit the distribution is not universal.
At larger Stokes numbers where caustics are
abundant (the formation of caustics is an activated process),
caustics and pair diffusion compete,
and the distribution assumes the universal shape summarised in \Eqnref{rho_asym_generaldim}.

\Eqnref{rho_asym_generaldim} determines
the moments $m_p$ of relative velocities for small separations.
As in the white-noise limit (see Ref.~\cite{Gus11b}) the moments
are found to be given by a sum of two contributions, a smooth contribution due to pair diffusion
and a singular contribution due to caustics. This form of the moments
is consistent with numerical results obtained in Ref.~\cite{Gus12} (see also \cite{Cen11}).
Our results also explain the scaling behaviours of relative particle-velocity structure
functions found in \cite{Bec10,Sal12}.

We remark that our results for the moments of relative velocities conditional on
a small separation $R$ is consistent with the $\st$-dependence of the average relative
velocity of inertial particles at small separations discussed by other authors. 
As first pointed out in Ref.~\cite{Abr75},
the average collision velocity of particles suspended in flows with a single scale scales as $\st^{-1/2}$
for large Stokes numbers.
The relative particle velocity (conditional on a small distance $R$) in turbulent flows with an inertial range scales as $\st^{1/2}$ for large Stokes numbers
\cite{Vol80,Wei93,Zai03,Meh07},
provided that $\st (\eta/\Lambda)^{2/3}\ll 1$.
Here $\eta$ is the Kolmogorov length of the flow, and $\Lambda$ is its integral length scale.

Our results have important implications for the collision rate of inertial particles, closely related to the first moment $m_1$ of relative velocities.
The universal form of $m_p$, \Eqnref{mp_generalform}, shows that the collision rate is a sum of two contributions (smooth and singular), rather than a product (see \Ssecref{structure_functions}).

Last but not least, in Sections~\secref{4} and~\secref{onedim} we compare to earlier results obtained in the white-noise limit \cite{Gus11b}.
We show that the solution found in \cite{Gus11b} is exact in the limit where the correlation length $\eta$ of the velocity
field tends to infinity.  In real systems $\eta$ is finite, and the power-law tails of the distribution of relative velocities are cut off.
By means of a series expansion in $\R/\eta$ (where $\R$ is the separation between two nearby particles) we compute
the far tail of the distribution of relative velocities in the white-noise limit.

\begin{figure}
\includegraphics[width=13.5cm]{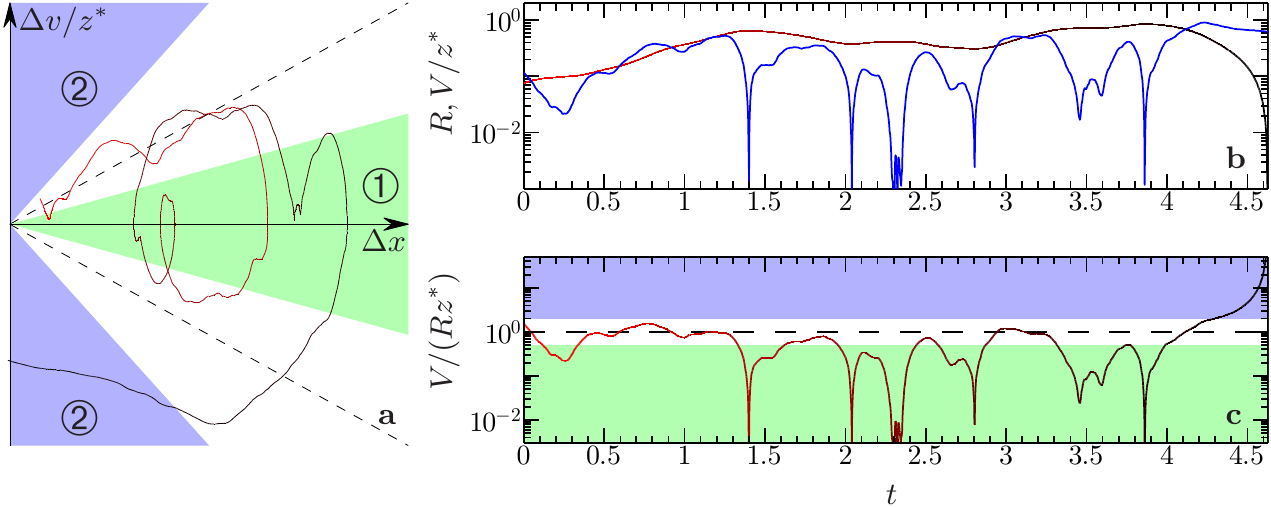}
\caption{
\label{fig:trajek}
{\bf a} Trajectory of separations and relative velocities of two close-by particles in one spatial dimension (solid line).
Time $t$ is colour coded: red at $t=0$ to black at large times.
Shaded regions illustrate the two asymptotic regimes $\V\ll\R$ (\ding{192}, light green) and $\V\gg\R$ (\ding{193}, light blue) separated by the matching curve $\V=\zstar\R$ (black dashed).
{\bf b} Distance $\R$ (red to black) and scaled relative speed $\V/\zstar$ (blue).
{\bf c} Relative magnitude of the quantities in {\bf b} (red to black).
Colored regions and black dashed as in {\bf a}.
Parameters: $\ku=1$ and $\st=5$ [$\epsilon^2=15$ according to \Eqnref{epsilon_def}].
Phase-space correlation dimension $D_2\approx 0.54$ and matching scale $\zstar\approx 1.8$.
}
\vspace{0.5cm}
\hspace{4.5cm}
\includegraphics[width=8.9cm]{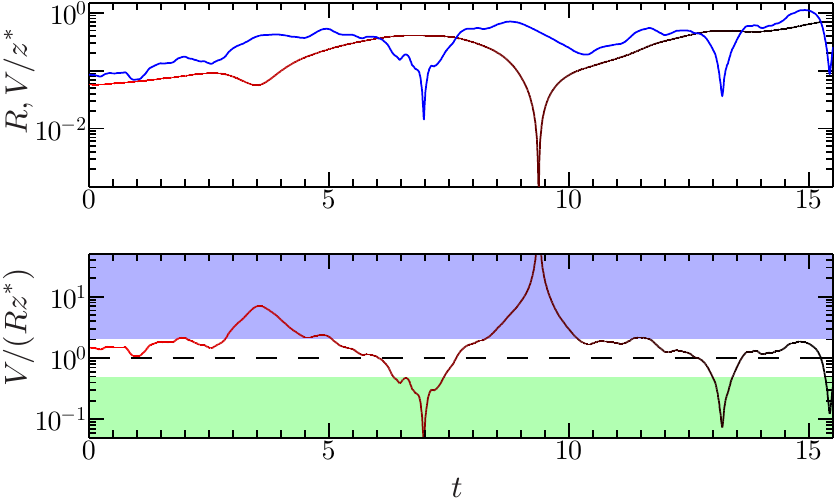}
\caption{
\figlab{trajek2d}
Same as \Figref{trajek} {\bf b} and {\bf c}, but in two spatial dimensions.
Parameters: $\ku=1$, $\st=1$ [$\epsilon^2=1/2$ according to \Eqnref{epsilon_def}].
Phase-space correlation dimension $D_2\approx 1.68$ and matching scale $\zstar\approx 0.38$.
}
\end{figure}

\begin{figure}
\includegraphics[width=13.5cm,clip]{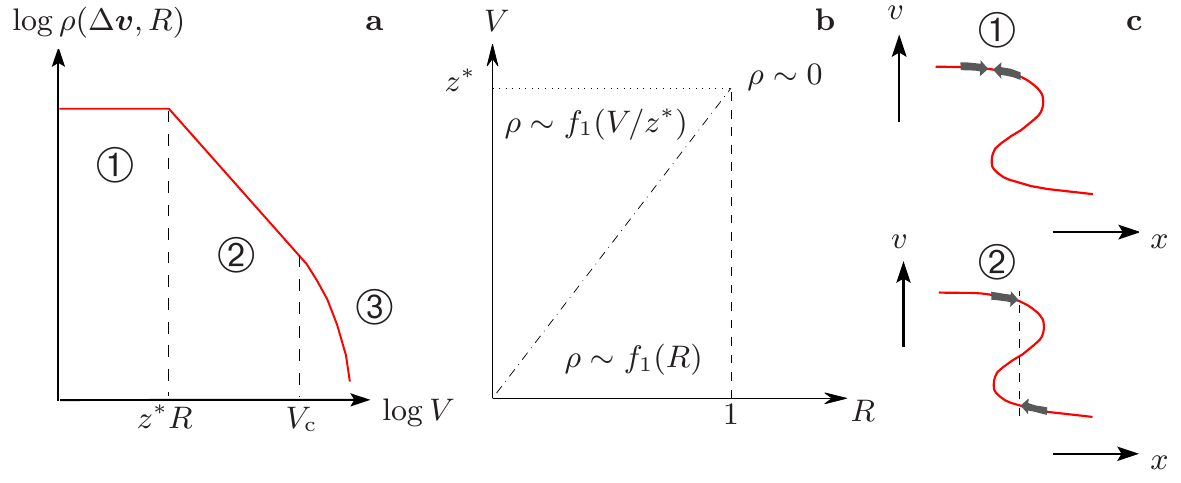}\\
\caption{
\label{fig:dist}
\label{fig:regions_universal}
\label{fig:1}
{\bf a} Schematic illustration of the distribution of relative velocities $\Delta\ve v$ as a function of $\V=|\Delta\ve v|$ at a given small distance $\R$, on a logarithmic scale.
{\bf b} Schematic plot of the matching regions (in dimensionless units) and the corresponding asymptotes
of the distribution $\rho(\Delta\ve v,\R)$ in \Eqnref{rho_asym_generaldim_ansatz}.
The matched distribution is continuous over the matching curve $\V=\zstar\R$ (dash-dotted) and the distribution is set to zero at the scales of the non-universal cut-offs $\Rc\sim 1$ and $\Vc\sim\zstar$.
{\bf c} Schematic illustration of different contributions to the distribution of relative velocities at small separations. The first case corresponds to particles approaching each other on the same branch of the phase-space manifold, the second case represents the singular caustic contribution: particles approach in position on different branches.
%  of the phase-space manifold.
}
\end{figure}

\section{Random velocity field}
\seclab{model}

In Section~\ref{sec:3} we employ general arguments to describe universal features of the distribution of separations and relative velocities.
To verify these arguments we use a \lq random-flow model'~\cite{Deu85,Wil07}, but we note that the validity of the results in
Section~\ref{sec:3} is not limited to this random-flow model.
In Sections~\secref{4} and~\secref{onedim} we consider the random-flow model in the \lq white-noise limit'.
In this limit we can explicitly calculate system-specific properties such as the phase-space correlation dimension necessary to parameterise the results in Section~\ref{sec:3}.

In our numerical simulations of Eq.~(\ref{eq:1}) we use a single-scale flow. It is convenient to use dimensionless variables.
We define $t' = \gamma t, \ve x' =  \ve x/\eta,
\ve v' = \ve v/(\eta\gamma)$, and $\ve u' = \ve u/(\eta\gamma)$ and drop the primes to simplify the notation.
In these dimensionless units Eq.~(\ref{eq:1}) takes the form
\begin{equation}
\label{eq:deutsch}
\dot{\ve x}=\ve v\,,\hspace{0.5cm}\dot{\ve v}=\ve u(\ve x,t)-\ve v\,.
\end{equation}
We write the random velocity field as $u=\ku\,\st\,\,\partial\phi/\partial x$ in one spatial dimension and as $\ve u=\ku\,\st\,\ve\nabla\phi\wedge\hat{\ve  e}_3/\sqrt{2}$  in two spatial dimensions. Here $\hat{\ve e}_3$ is the unit vector in the $z$-direction. We assume $\phi$ to be homogeneous in space and time, with mean and correlation function
\begin{equation}
\langle \phi(x,t)\rangle = 0\,,\quad\mbox{and}\quad \langle \phi(x,t)\phi(x',t)\rangle=C(|x'-x|,|t'-t|)\,.
\eqnlab{model_corrfun}
\end{equation}
Angular brackets denote averages over an ensemble or realisations of the velocity field.
The velocity field is assumed to be locally smooth, implying $C(x,t)-C(0,t) \propto x^2$ for small values of $x$.
Furthermore, the correlation function $C(|x|,|t|)$ is assumed to be normalised such that $C(0,0)=1$ and to decay towards zero for large values of $|x|$ and $|t|$.

As pointed out above, the dynamics of (\ref{eq:1}) is determined by the two dimensionless parameters $\st$ and $\ku$.
The Stokes number $\st=(\gamma \tau)^{-1}$ is a dimensionless measure
of the damping of the particle velocities relative to the flow. In the overdamped limit, $\st\rightarrow 0$,
the particles are advected by the flow. This limit is well understood, see for example
\cite{Fal01} for a summary of what is known. In the underdamped limit $\st\rightarrow \infty$, by contrast, the particles
form random gas, and their relative velocities are described by gas kinetics~\cite{Abr75}.

The Kubo number $\rm Ku$ is determined by the typical length and time scales of the flow.
The fluctuations of the random velocity field are characterised by a single
spatial scale (the correlation length $\eta$) and a single time scale (the correlation time $\tau$).
Gustavsson {\em et al.} \cite{Gus08b} discuss relative velocities at high Stokes numbers in flows with a range of scales (fully
developed turbulent flows for example). We comment in \Ssecref{moments_singular} on how the results described
in this paper connect to those of Gustavsson {\em et al.} \cite{Gus08b}.
In terms of the typical size $u_0$ of the flow velocity, the Kubo number is
given by $\ku= u_0 \tau/\eta$. In turbulent flows, $\ku$ is of order unity.
The results in Section~\ref{sec:3} are valid for general values of $\ku$ and $\st$ provided that caustics occur. In this case the results give the dominant contribution to the distribution of separations and relative velocities.
The rate of caustic formation increases as $\st$ increases, implying that the results in Section~\ref{sec:3} are expected to become more accurate for larger values of $\st$.

The results in Sections \secref{4} and \secref{onedim} are obtained in the white-noise limit $\ku\rightarrow 0$ and $\st\rightarrow \infty$ such that
\begin{align}
\epsilon^2\equiv e_d\ku^2 \st
\eqnlab{epsilon_def}
\end{align}
remains constant. The precise value of the constant $e_d$ is a matter of 
convention. In this article, $e_d$ is defined so that $\epsilon^2$ is identical to ${\cal D}$ in Ref.~\cite{Gus08}.
In one spatial dimension the coefficient is $e_1=3$, and in incompressible two-dimensional flows the coefficient is $e_2=1/2$.

In the limit  $\ku\rightarrow 0$ and $\st\rightarrow \infty$ such that
$\ku^2 \st$ remains constant the particles experience the flow
as a white-noise signal, and their dynamics is \lq ergodic\rq{}: a given particle
uniformly samples configuration space, and the instantaneous
configuration of the flow field is irrelevant to the long-term fluctuations of the particle trajectory.

\section{Distribution of relative velocities and separations}
\label{sec:3}

In this section we show how to compute the joint distribution of relative velocities
and separations for particles suspended in mixing flows.
% The resulting distribution is, independently of the nature of the flow, valid in certain asymptotic limits of relative velocities and separations provided that the
% relative velocities are not too large and that spatial separations are sufficiently small.
Throughout this section we assume that the spatial separation between the two particles in question is much smaller than the correlation length of
the flow ($\R\ll 1$ in dimensionless units). This is the case relevant for collision velocities of small particles.
We consider the dynamics of a pair of particles with spatial separation vector $\Delta\ve x = \ve x_2-\ve x_1$
and relative velocity $\Delta\ve v=\ve v_2-\ve v_1$ to find the joint distribution $\rho(\Delta\ve v,\Delta\ve x)$.
The relative speed is denoted by $\V = |\Delta\ve v|$, and the distance by $\R = |\Delta\ve x|$.

\subsection{Matching asymptotic limits of the distribution}
Fig.~\ref{fig:trajek}{\bf a} shows a trajectory of a pair of particles
exploring the space of separations $\Delta x$ and relative velocities $\Delta v$ in one spatial dimension.
\Figref{trajek}{\bf b} shows $\R$ and $\V$, and \Figref{trajek}{\bf c} shows the magnitude of $\V/\R$ as a function of time for the trajectory in \Figref{trajek}{\bf a}.
The parameter $\zstar$ in \Figref{trajek} represents typical values of the relative magnitude of speed and distance between particles, $\zstar\approx\overline{V/\R}$, where $\overline{X}$ denotes the time average of $X$.
%
% For d=1 with positive maximal Lyapunov exponent particles 
% have a vanishing probability to be at $\R=0$, 
% which allow finite values of $\overline{V/\R}$.
% If d>1 it is quite clear that the geometric prefactor allows integrability.
%
The distribution of $\Delta x$ and $\Delta v$ is determined by the following observations.
\begin{enumerate}
\item Large relative velocities cause large changes in small particle separations. Consider
a particle pair with $V\gg R$ (in dimensionless variables). As the particles
move, $R$ changes rapidly while $V$ remains relatively unaffected by the motion
at time scales smaller than $\st$.
A trajectory exhibiting this behaviour is shown
in Fig.~\ref{fig:trajek}. The region $V\gg R$
is labeled $2$ in Fig.~\ref{fig:trajek}{\bf a} (also shown in \Figref{trajek}{\bf c}).
For a given large value of $V$, we find that all values of $R$ such that $V\gg R$ are equally probable.
In other words we expect $\rho(\Delta v,\Delta x)$ to be approximately independent of $R$ in region $2$.

\item
When by contrast $V\ll R$, then the separation $R$ does not change much,
whereas the relative speed $V$ is found to fluctuate greatly.
This is shown in Fig.~\ref{fig:trajek} (region $1$). In this
region, we expect the distribution to be roughly independent of $V$.

\item The trajectory in Fig.~\ref{fig:trajek} spends an appreciable fraction
near the boundary $R \approx V$ between regions $1$ and $2$.
Starting from region $1$, the particle pair may eventually attain  values of relative speed $V$
comparable to the distance $R$. In this case it may wander inwards, entering region $1$.
\end{enumerate}

The nature of the trajectory shown in Fig.~\ref{fig:trajek} is reflected strongly in
the distribution of relative velocities at small separations, shown in Fig.~\ref{fig:dist}{\bf a}
(schematically for a fixed, small value of $R$).
Region $1$ in Fig.~\ref{fig:trajek} gives rise to the body of the distribution, the plateau
at small values of $V$ (as mentioned above, the distribution is
expected to be approximately independent of $V$ in this region). As we show below, region $2$ gives rise to
the power-law shown in Fig.~\ref{fig:dist}{\bf a}. In this regime the distribution
is approximately independent of $R$. 

What is the relation between these observations
and the fact that phase-space manifolds fold when caustics form?
Consider a trajectory approaching a finite value of $V$ as $R\rightarrow 0$
(region $2$ in Fig.~\ref{fig:trajek}). As $\R\rightarrow 0$,
the relative speed $V$ can remain finite only if the particles approach on two different branches of
the manifold (as shown schematically in Fig.~\ref{fig:1}{\bf c}).
The power-law part of the distribution in Fig.~\ref{fig:dist}{\bf a}
is caused by spatial clustering at small separations in combination with caustics.
When caustics form, they project the distribution at $\V\sim\zstar\R$ (if $\R\ll 1$) towards smaller separations with approximately constant $\V$.
This effectively mirrors the distribution in the line $\V\sim\zstar\R$, giving rise to power laws in the distribution of $\V$ for small separations.
The body of the distribution at small relative velocities, by contrast, corresponds to particles
approaching on the same branch (or close-by branches).
The far tails in Fig.~\ref{fig:dist}{\bf a} (labeled $3$ in \Figref{dist}{\bf a}) are discussed at the end of this Section and in \Ssecref{moments_singular}.

In order to explain the observations described above and to extend them to higher spatial dimensions,
we consider
the dynamics of $\Delta\ve v$ and $\Delta\ve x$. It is determined by linearising \eqnref{deutsch}
\begin{equation}
\label{eq:deutsch_sep}
\frac{\rm d}{{\rm d}t} {\Delta\ve x}=\Delta\ve v\,,\hspace{0.5cm}
\frac{\rm d}{{\rm d}t}{\Delta\ve v}=\Delta\ve u(\ve x,\Delta\ve x,t)-\Delta\ve v\,.
\end{equation} 
Here $\Delta\ve u(\ve x,\Delta\ve x,t)\equiv\ve u(\ve x+\Delta\ve x,t)-\ve u(\ve x,t)$.
We study the limit of $R\ll 1$ so that $\Delta\ve u\approx \ma A\Delta\ve x$,
where $\ma A$ is the matrix of fluid gradients, with elements $A_{ij}=\partial u_i/\partial x_j$.

In one spatial dimension (region $2$ in Fig.~\ref{fig:trajek}),
particle pairs approach small separations at large relative velocities on different
branches of the multi-valued velocity field, $\V\gg\R$.
In higher spatial dimensions too the phase-space manifold may fold over (in two spatial
dimensions this is illustrated in Fig.~2 in \cite{Gus13}).
This allows the trajectory to visit the region $\V \gg \R$ also in higher spatial dimensions, as is shown in \Figref{trajek2d}.
Particles rush past each other with velocities from different branches of
the multi-valued velocity field and the distribution of particle separations becomes uniform at small separations.
When $\V\gg \R$, the driving $\Delta\ve u\approx\ma A\Delta\ve x$ in Eq.~(\ref{eq:deutsch_sep}) is negligible,
and in region $2$ the dynamics in $\Delta\ve x$ and $\Delta\ve v$ is approximated by
\begin{equation}
\Delta\ve v_t \approx \Delta\ve v_0 + (\Delta\ve x_0-\Delta\ve x_t)\approx \Delta\ve v_0\,,
\eqnlab{region2_approx}
\end{equation}
Here we assume $|\Delta\ve v_0|\gg|\Delta x_0|$ and the approximation is valid until $|\Delta\ve x_t|$ becomes comparable to $|\Delta\ve v_0|$.
In this case the separation $\Delta\ve x$ changes with approximately constant velocity, $\Delta\ve x=\Delta\ve x_0+\Delta\ve v_0t$.
This is clearly seen in one spatial dimension
in Fig.~\ref{fig:trajek}{\bf a}; the trajectory is predominantly horizontal in region $2$,
particles approach in position on two different branches of the folded phase-space attractor.
A corresponding path in two spatial dimensions that goes through region 2 is shown as a function of time in \Figref{trajek2d}.
Such paths represent the \lq singular caustic contribution' to $\rho(\Delta\ve v, \Delta\ve x)$,
and give rise to large moments of $\V$ at small values of $\R$.
For a given value of $\Delta\ve v_0$, the fact that the motion is uniform implies that all separations $\Delta\ve x$
such that $|\Delta\ve v_0|\gg\R$ are equally likely, and thus the distribution $\rho(\Delta\ve v,\Delta\ve x)$
is expected to be independent of $\Delta\ve x$.
Further, it follows from  isotropy that all directions of relative incoming velocities are equally likely.
This implies that $\rho(\Delta\ve v,\Delta\ve x)$ is a function of $V$ only.
We conclude:
\begin{equation}
\rho(\Delta\ve v,\Delta\ve x)=\fII(\V)\mbox{ for } \V\gg\R\,,
\label{rho1d_VggX}
\end{equation}
where $\fII(\V)$ is a function to be determined.
We emphasise that $\fII(\V)$ must be a universal asymptote (at values of $\R$ much smaller than both unity and $\V$) because in this limit the dynamics is insensitive to the nature of the stochastic driving.

Region $1$ in Figs.~\ref{fig:trajek} and~\ref{fig:trajek2d} corresponds to
the condition $\V \ll \R$. In this case, the dynamics
of  $\Delta\ve x$ and $\Delta\ve v$ is approximately given by:
\begin{equation}
\label{eq:OU}
\frac{\rm d}{{\rm d}t} \Delta\ve x \approx 0\,,\quad\mbox{and}\quad
\frac{\rm d}{{\rm d}t} \Delta\ve v  = \Delta\ve u -\Delta\ve v \,.
\end{equation}
This limit describes particle pairs moving at a constant spatial separation for a long time, because their relative velocity is small.
This corresponds to the fluctuating vertical trajectories in Fig.~\ref{fig:trajek}{\bf a}.
An example of a two-dimensional trajectory which passes through region 1 is shown in \Figref{trajek2d}.
Such paths arise in systems with single-valued, smooth particle-velocity fields, but cannot bring point particles in contact
(cannot achieve arbitrarily small values of $\R$ in finite time).
We term this contribution the \lq smooth contribution' to $\rho(\Delta\ve v, \Delta\ve x)$.
The dynamics of $\Delta\ve v$ in this limit depends on the fluctuations of $\Delta\ve u(\ve x,\Delta\ve x,t)$.
It is thus not universal.
At constant separations, $\Delta\ve x=\Delta\ve x_0$, the relative fluid velocity $\Delta\ve u(\ve x,\Delta\ve x_0,t)$
is weakly coupled to the relative motion between particles through non-ergodic effects.
In the case of white-noise flows studied in Sections~\secref{4} and~\secref{onedim}, the $\Delta\ve v$-equation becomes
an Ornstein-Uhlenbeck equation. Its steady-state solution  is a Gaussian in $\Delta\ve v$ with a $\Delta\ve x$-dependent variance.
% In this section, to keep the discussion general, we consider the limit of $\V\to 0$
Universality emerges in the limit of $\V\to 0$, where the distribution $\rho(\Delta\ve v,\Delta\ve x)$ approaches a
function of $\Delta\ve x$ only.
% This behaviour is universal.
It follows from isotropy that this function can only be a function of $\R$. We conclude:
\begin{equation}
\rho(\Delta\ve v,\Delta\ve x)=\fI(\R)\mbox{ for } \V\ll\R\,,
\label{eq:rho_VllX}
\end{equation}
where $\fI(\R)$ is a function to be determined.
The functions $\fI$ and $\fII$ are found by invoking the following two
principles:
\begin{enumerate}
\item
\label{item:principle1}
Eqs.~(\ref{eq:rho_VllX}) and (\ref{rho1d_VggX}) are matched
in the $\R$-$\V$-plane along the curve $\V=\zstar\R$.
The scale factor $\zstar$ is determined by the relative importance of the two regions.
This is an approximation, because the expressions that are matched are asymptotic (valid in regions $1$ and $2$ of Figs.~\ref{fig:trajek} and~\ref{fig:trajek2d}), and because it is assumed that the boundary between regions $1$ and $2$ can
be parameterised by $\V=\zstar\R$.
In the limit of small values of $\R$ studied here, we motivate the second approximation by the fact that the
change in time of $\Delta\ve v/\R$ only depends on the quotient $\Delta\ve v/\R$, and not on $\Delta\ve v$ and $\R$ separately when $\R$ is small.
We construct the approximate distribution to be rotationally symmetric in both $\Delta\ve x$ and $\Delta\ve v$.
In this case the distribution depends upon $\R$ and $\V$ only, and the dynamics is constrained to a single independent variable $z=\V/\R$.
We denote the scale that distinguishes small from large values of $z$ by a constant $\zstar\approx\overline{\V/\R}$ (Figs.~\figref{trajek}{\bf c} and \figref{trajek2d}{\bf c} show that typical values of $\V/\R$ are close to $\zstar$).
This yields the matching curve $\V=\zstar\R$.
We note that in spatial dimensions larger than one, the dynamics of $z$ is coupled to that of $z_\R\equiv\Vr/\R$,
where $\Vr$ is the radial relative velocity. This coupling indicates that an improved matching curve may break the rotational symmetry in $\Delta\ve v$.
In this paper we do not consider this complication.
\item
\label{item:principle2}
When the maximal Lyapunov exponent of (\ref{eq:1}) is positive (as is the case for incompressible flows \cite{Bec03b,Dun05,Gus11} and for compressible flows for large enough 
values of $\st$, see Ref.~\cite{Meh05}) and provided that the system size is finite, then the phase-space manifold forms a fractal attractor
(an example is shown in \Figref{attractor}).
We characterise the fractal clustering in phase-space by the \lq phase-space correlation dimension' $D_2$.
Consider the distribution $P(w)$ of small phase-space separations $w\equiv\sqrt{\R^2+(V/\zstar)^2}$.
The phase-space correlation dimension $D_2$ is defined by the form of the distribution $P(w)$ as $w\rightarrow 0$:
\begin{equation}
P(w)\sim w^{D_2-1} \quad\mbox{as $w\rightarrow 0$}\,.
\eqnlab{ws}
\end{equation}
The phase-space correlation dimension $D_2$ can take values up to $2d$ (where $d$ is the spatial dimension).
We determine the functions $\fI$ and $\fII$ from the scaling behaviour in (\ref{eq:ws}) for small values of $w$.
\end{enumerate}
Invoking the first principle, we match Eqs.~(\ref{eq:rho_VllX}) and (\ref{rho1d_VggX}) at $\V=\zstar\R$ with a cut-off $\Vc$ at $\R=1$.
By continuity we must have $\Vc=\zstar$ and $\fII(y)=\fI(y/\zstar)$ for any value of $y$.
This yields:
\begin{align}
\rho(\Delta\ve v,\R)\sim
\R^{d-1}
\left\{
\begin{array}{lll}
\fI(\R) & \mbox{for }\V\le \zstar\R\,,\,\,\R\le 1\mbox{ and }\V\le\zstar & \I\cr
\fI(\V/\zstar) & \mbox{for }\V>\zstar\R\,,\,\,\R\le 1\mbox{ and }\V\le\zstar & \II\cr
0 & \mbox{for }\R > 1\mbox{ or }\V>\Vc = \zstar & \III\cr
\end{array}
\right.\,.
\eqnlab{rho_asym_generaldim_ansatz}
\end{align}
The factor $\R^{d-1}$ in Eq.~(\ref{eq:rho_asym_generaldim_ansatz}) is simply the geometric factor for an isotropic distribution in a
spherical coordinate system of spatial separations.
The relation between the different regions referred to in 
\Eqnref{rho_asym_generaldim_ansatz} is
illustrated in \Figref{regions_universal}{\bf b}.

In region $2$ (or for large enough separations in region $3$) the spatial part of the distribution is uniform and the $\R$-dependence is given by $\R^{d-1}$.
In region $1$, by contrast, fractal spatial clustering modifies this behaviour.

Invoking the second principle, we determine $\fI$ by applying
the condition (\ref{eq:ws}) to \Eqnref{rho_asym_generaldim_ansatz}.
We change variables from $\R$ to $w$ and introduce a second independent set of spherical coordinates in $\Delta\ve v$. We integrate the angular coordinates away to find
\begin{align}
P(w)&\sim\int_0^{\zstar w}{\rm d}\V\V^{d-1}\left.\rho(\Delta\ve v,\R)\frac{\partial\R}{\partial w}\right|_{\R=\sqrt{w^2-(\V/\zstar)^2}}\,.
\end{align}
Here we have used that $\V$ is bounded by $\zstar w$.
Using $\rho(\Delta\ve v,\R)$ from \eqnref{rho_asym_generaldim_ansatz} and changing the integration variable
to $\V=w\zstar\sin(\gamma)$ with $0\le\gamma<\pi/2$  we find:
\begin{align}
P(w)&\sim
2w^{2d-1}\zstar^{d}\int_0^{\pi/4}{\rm d}\gamma (\sin(\gamma)\cos(\gamma))^{d-1}\fI(w\cos(\gamma))\,.
\end{align}
Comparing this expression to Eq.~(\ref{eq:ws}) we find the form of $\fI$: $\fI(y)\sim y^{D_2-2d}$.
Inserting this power law into \eqnref{rho_asym_generaldim_ansatz} we obtain:
\begin{align}
\rho(\Delta\ve v,\R)\sim
\R^{d-1}
\left\{
\begin{array}{lll}
\R^{D_2-2d} & \mbox{for }\V\le \zstar\R\,,\,\,\R\le 1\mbox{ and }\V\le\zstar & \I\cr
|\Delta\ve v/\zstar|^{D_2-2d} & \mbox{for }\V>\zstar\R\,,\,\,\R\le 1\mbox{ and }\V\le\zstar & \II\cr
0 & \mbox{for }\R > 1\mbox{ or }\V>\zstar & \III\cr
\end{array}
\right.\,.
\eqnlab{rho_asym_generaldim}
\end{align}
\begin{figure}
\includegraphics[width=6.5cm]{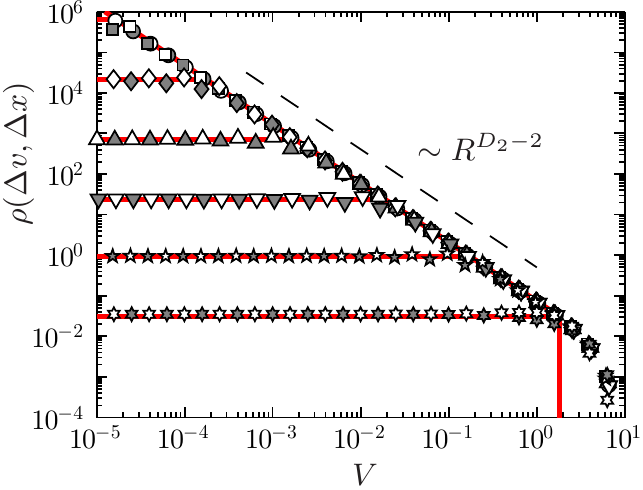}
\caption{\label{fig:distribution_1d_Ku1}
Distribution $\rho(\Delta v,\Delta x)$ in one spatial dimension as a function of $\Delta v$ for $\Delta x=10^{-6}$ ($\circ$), $10^{-5}$ ($\square$), $10^{-4}$ ($\diamond$), $10^{-3}$ ($\vartriangle$), $0.01$ ($\triangledown$), $0.1$ ($\star$) and $1$ ($\ast$).
Symbols show results from numerical simulations of the model described in \Secref{model}.
White/shaded symbols show data for positive/negative values of $\Delta v$, respectively.
Solid red lines show the asymptotic distribution \eqnref{rho_asym_generaldim}.
Same parameters as in \Figref{trajek}.
}
\vspace{0.5cm}
\includegraphics[width=13.15cm]{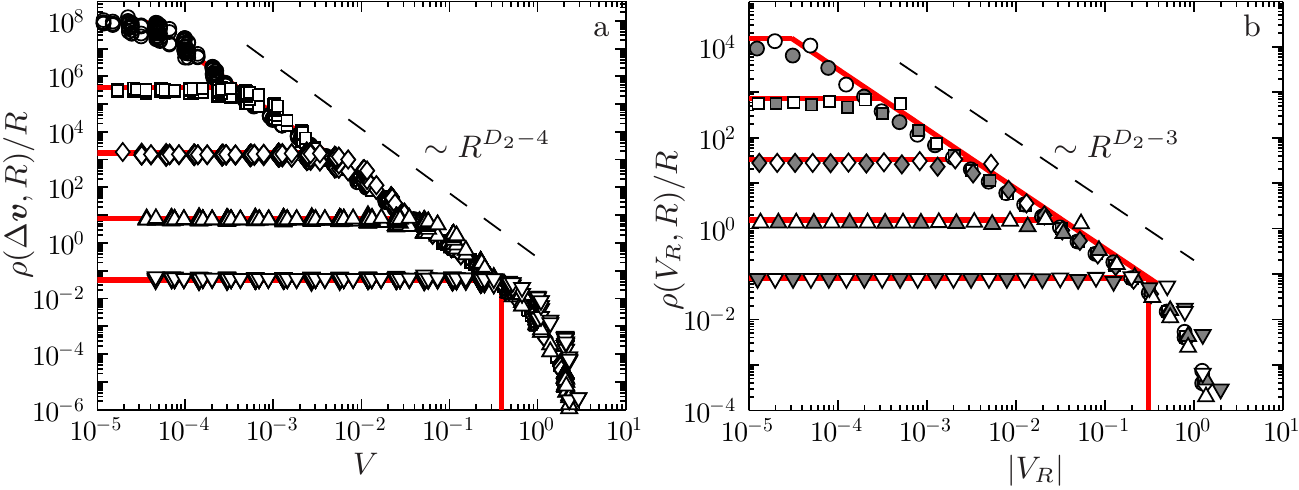}
\caption{\label{fig:distribution_2d_Ku1}
{\bf a} Distribution $\rho(\Delta\ve v,\R)/\R$ in two spatial dimensions as a function of $\V$ for $\R=10^{-4}$ ($\circ$), $10^{-3}$ ($\square$), $0.01$ ($\diamond$), $0.1$ ($\vartriangle$) and $1$ ($\triangledown$).
For each value of $\R$, $\rho(\Delta\ve v,\Delta\ve x)/\R$ is shown for all values of $\Delta\ve v$ as a function of $\V = |\Delta \ve v|$.  
Otherwise same notation as in \Figref{distribution_1d_Ku1}.
{\bf b} Same but for $\rho(\Vr,\R)/\R$.
Theory according to \eqnref{rhoVr_asym2_generaldim}.
Same parameters as in \Figref{trajek2d}.
}
\end{figure}
The asymptotic result \eqnref{rho_asym_generaldim} is universal, it describes the joint distribution of relative velocities and separations
of particles suspended in randomly mixing or turbulent flows. It does not depend on the particular form
of the fluctuations of the flow velocity. Eq.™(\ref{eq:rho_asym_generaldim}) is valid as long as
there is fractal phase-space clustering and as long as caustics occur, so that the region $\V\gg\R$ is accessible.

Eq.~(\ref{eq:rho_asym_generaldim})  is the main result of this paper.
We use it to derive power-law forms of the distribution of $\Vr$ and $\R$,  as well as of the power-law scalings
of the moments of relative velocity at finite Kubo numbers.
Our result \eqnref{rho_asym_generaldim} is consistent with the qualitative observations summarised in the beginning
of this section.

The form of the distribution of relative velocities at small separations  $\R$ is illustrated schematically  in Fig.~\ref{fig:dist}{\bf a}.
First, when $\V\ll\zstar\R$, then \Eqnref{rho_asym_generaldim}
predicts that at small separations the distribution of $\Delta\ve v$
becomes approximately independent of $\Delta\ve v$, and that its amplitude
scales as $\R^{D_2-d-1}$.
This is region $1$ in Figs.~\figref{trajek}--\figref{dist}. At larger values of relative velocities,
$\V\gg\zstar\R$, the distribution of relative velocities  exhibits a power law $\rho(\Delta\ve v,\R)\sim\R^{d-1}\V^{D_2-2d}$.
This power law (in region $2$) reflects the existence of caustics
giving rise to large relative velocities at small separations.
We observe that the algebraic decay is so slow
that the moments $m_p(\R) \equiv \langle|\Vr|^p\rangle_{\R}$
would diverge for $p \ge d-D_2$ if the algebraic tails were not cut off.
This cut off, $\Vc\sim\zstar$, arises simply because the magnitude of the
relative velocities cannot exceed the largest magnitude of the relative driving force $\Delta\ve u$.
The region $\V\gg  \Vc$ is referred to as region $3$ in Fig.~\ref{fig:dist}.
The behaviour in this region is not universal.
For particles suspended in a flow with a range of spatial scales
we expect that the tail of the distribution
of relative velocities is determined by the variable-range
projection principle (derived in \cite{Gus08},
see also Ref.~\cite{Pan13}). 
This point is further discussed in \Ssecref{moments_singular}.
In \Secref{onedim} we calculate the cut-off explicitly in a one-dimensional white-noise model.
The simplest approximation is to just set $\rho=0$ for $V>\Vc$, as in Eq.~(\ref{eq:rho_asym_generaldim}).

The result  \eqnref{rho_asym_generaldim} is compared to results of numerical simulations of the random-flow model for
$\ku\sim 1$ and $\st\sim 1$ in Figs.~\figref{distribution_1d_Ku1} and~\figref{distribution_2d_Ku1}.
% (we had to take $\st>1$ in the one-dimensional case because choosing $\st=1$ would give a negative maximal Lyapunov exponent).
We find that \eqnref{rho_asym_generaldim} describes the numerical results well in the region
where it is expected to apply, namely for
$\R\ll 1$ and $\V\ll\zstar$.
It should be noted that for smaller values of $\st$, caustics occur less frequently and the relative weight of the power-law tails to the body of the distribution becomes smaller.

Closer inspection of the numerical results reveals a number of system-specific properties that are not described by \eqnref{rho_asym_generaldim}.
First, the distribution at $\V\sim\zstar\R$ is asymmetric in $\Delta\ve v\to-\Delta\ve v$ if $\R>0$.
Second, at large spatial separations, $\R>1$, the detailed form of the distribution is system specific.
The large-$R$ behaviour for particles suspended in a single-scale flow is expected to be different
from the large-$R$ behaviour for particles suspended in a multi-scale flow (see \Ssecref{moments_singular}).
Third, when $\V\gg\Vc$ we expect that the distribution is  cut off,  corresponding to the cut off at large values of $\R$.
This cut off is due to the fact that the driving force is unlikely to obtain values much larger than some typical value determined by the cut-off in $\R$.
Note however that if the dissipative range (where $\langle |\Delta \ve u|^2\rangle \sim R^2$) extended to infinity
then infinitely large relative velocities could in principle occur, and the $\V$-tails would not be cut off.
This is the case in the limit $\eta\rightarrow\infty$.
In \Ssecref{moments_singular} we show how to compute the large-$\V$ cutoff for general correlation functions of $\Delta\ve u$.
We find that the decay of $\Delta\ve v$ for $\V\gg\Vc$ is faster than algebraic for single-scale flows, or flows with an inertial range.
In \Secref{onedim} we show how to calculate the system-dependent properties of the distribution for a one-dimensional white-noise model.

\subsection{Distribution of collision velocities}
\sseclab{DistCollVel}

We now derive universal properties of the distribution of radial relative velocities from \eqnref{rho_asym_generaldim}.
Radial relative velocities determine the collision rate between the suspended particles.
Universal behaviour is expected far from the matching boundaries, that is when  $\R\zstar$ and $|\Vr|$ are sufficiently different from each other, and  when
$|\Vr|\ll\zstar$ as well as $\R\ll 1$. We project $\Delta\ve v$ onto the unit vectors of the time-dependent spherical coordinate system
$\hat{\ve e}_R,\hat{\ve e}_{\phi_1},\dots,\hat{\ve e}_{\phi_{d-1}}$ ($\hat{\ve e}_R$ is aligned with $\Delta \ve x$ at all times):
\begin{equation}
V_\alpha\equiv\Delta\ve v\cdot\hat{\ve e}_\alpha\,,
\eqnlab{Valpha}
\end{equation}
with $\alpha=\R,\phi_1,\dots,\phi_{d-1}$.
Integrating on the projections $V_{\phi_i}$ with $i=1,\dots,d-1$ and inspecting the limiting behaviour of the result
$\rho(\Vr,\R)$ when $|\Vr|$ is much larger or smaller than $\zstar\R$, we find the following asymptotic distribution for $\R$ and $\Vr$
\begin{align}
\rho(\Vr,\R)
&\sim
\R^{d-1}
\left\{
\begin{array}{lll}
\R^{D_2-d-1} & \mbox{for }|\Vr|<\zstarx{\R}\R\,,\,\,\R\le 1\mbox{ and }|\Vr|\le\zstar & \I\cr
|\Vr/\zstar|^{D_2-d-1} & \mbox{for }|\Vr|\ge\zstarx{\R}\R\,,\,\,\R\le 1\mbox{ and }|\Vr|\le\zstar & \II\cr
0 & \mbox{for }\R > 1\mbox{ or }|\Vr|>\zstar & \III\cr
\end{array}
\right.
\eqnlab{rhoVr_asym2_generaldim}
\end{align}
provided that $D_2<d+1$. Here the matching scale $\zstarx{\R}$ is determined from the exact integration of \eqnref{rho_asym_generaldim}.
We find:
\begin{align}
\zstarx{\R}=\zstar\left(\frac{D_2-2d}{D_2-d-1}\frac{\Gamma((2d-D_2)/2)}{\Gamma((d-D_2+1)/2)\Gamma((d+1)/2)}\right)^{1/(D_2-d-1)}\,.
\end{align}
If, by contrast, $D_2>d+1$, then the distribution is uniform, $\rho(\Vr,\R)\sim\R^{d-1}$.
The result  \eqnref{rhoVr_asym2_generaldim} is compared to data from numerical simulations for $\ku=\st=1$ in \Figref{distribution_2d_Ku1}. We observe good agreement.

\subsection{Moments of collision  velocities}
\sseclab{moments}

Eq.~(\ref{eq:rhoVr_asym2_generaldim}) determines the moments of the collision velocity.
We define the moments of the radial relative velocity as
\begin{equation}
\label{eq:momentsmp}
m_p(\R)=\int\!{\rm d}V_R\, |V_R|^p\, \rho(V_R,R)
\equiv\langle|\Vr|^p\rangle\,.
\end{equation}
Multiplying  \eqnref{rho_asym_generaldim} with $|\Vr|^p$ and
integrating on  $\Delta\ve v$  we find:
\begin{align}
m_p(\R\ll 1)\sim b_p\R^{p+D_2-1}+c_p\R^{d-1}\,.
\eqnlab{mp_generalform}
\end{align}
The constants $b_p$ and $c_p$ depend on both the power-law body and on the tail of the distribution of relative velocities at small separations.
The tail give rise to a non-universal contribution to $b_p$ and $c_p$ for large values of $p$.
At small separations, the tails are independent of $R$, thus they cannot change the power laws in
\eqnref{mp_generalform}. It follows that the power laws in Eq.~(\ref{eq:mp_generalform}) are universal.
In order to calculate the coefficients $b_p$ and $c_p$, the tails of the distribution must
be properly accounted for. We show in \Secref{onedim} how this can be done for
a one-dimensional model in the white-noise limit.

The first term in Eq.~(\ref{eq:mp_generalform}) results predominantly from the smooth contribution to $\rho(V_R,R)$,
corresponding to region $1$ in Fig.~\ref{fig:dist}. The second term
represents the singular contribution due to caustics, it
corresponds to regions $2$ and $3$ in Fig.~\ref{fig:dist}.
The factor $\R^{d-1}$ multiplying the caustic contribution comes from the fact that the spatial part of the distribution is uniform in regions $2$ and $3$ (for large enough $\V$), or equivalently because only caustics that project the particles in a particle pair towards each other with a small enough relative angle contribute
 to the moment at small values of $\R$~\cite{Gus12}.
For $p=1$, \Eqnref{mp_generalform} corresponds to an expression for the average relative velocities
determining the collision rate. It is consistent
with the form proposed by Wilkinson {\em et al.} \cite{Wil06}. Eq.~(7) in that paper
suggests that the collision rate is the sum of two terms,
a smooth contribution (corresponding to advective collisions)
and a singular term (corresponding to collisions due to caustics).
Fractal clustering was not considered in Ref.~\cite{Wil06}.
% corresponding to $D_2 \approx d$ in Eq.~(\ref{eq:mp_generalform}).

\begin{figure}
\includegraphics[width=6.5cm]{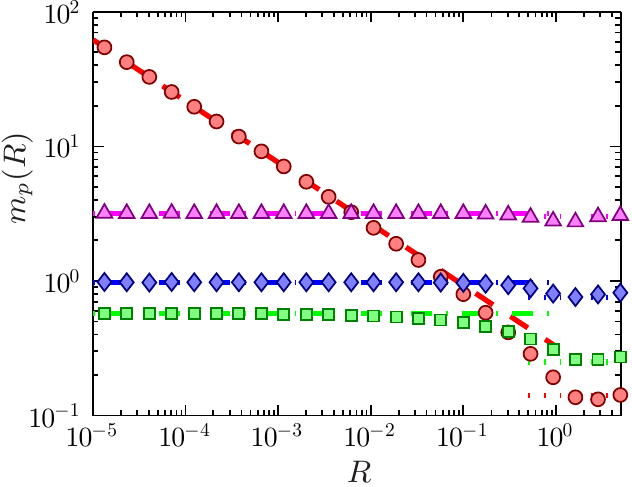}
\caption{\label{fig:moments_1d_Ku1}
The velocity moments $m_p(\R)$ in one spatial dimension as a function of 
$\R$ for  $p=0$ (red,$\circ$), $p=1$ (green,$\square$), $p=2$ (blue,$\diamond$) and $p=3$ (magenta,$\vartriangle$).
Same parameters and numerical data as in \Figref{distribution_1d_Ku1}.
Dash-dotted lines show first term in \eqnref{mp_generalform} with fitted $c_p$.
Dashed lines show second term in \eqnref{mp_generalform} with fitted $b_p$ (negative for $p>1$).
Dotted lines show large-$\R$ asymptotics with fitted prefactor.
}
\vspace{0.5cm}
\includegraphics[width=13.5cm]{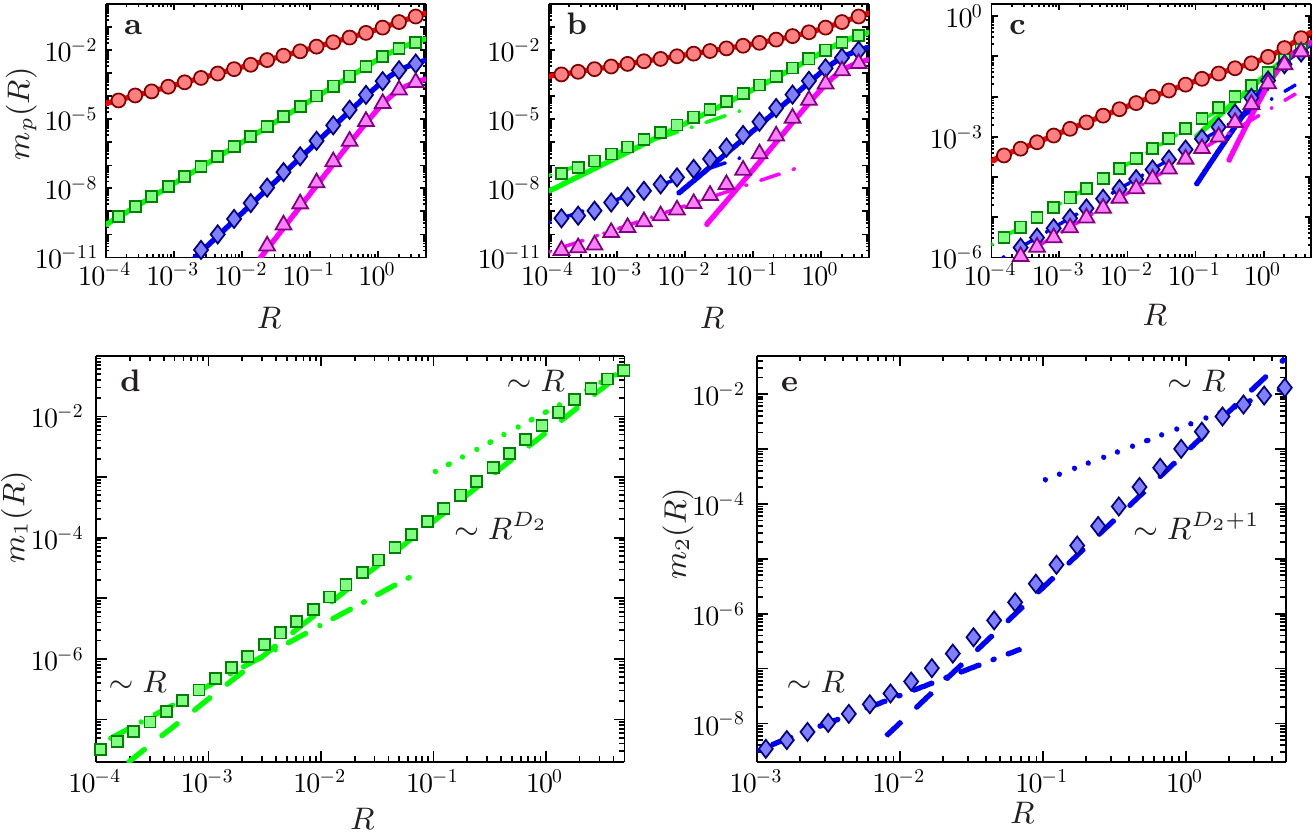}
\caption{\label{fig:moments_2d_Ku1}
Panels {\bf a}--{\bf c}:
Same as \Figref{moments_1d_Ku1} for two spatial dimensions.
Solid lines show the smooth contribution for general values of $\R$ according to \eqnref{mp_smooth},
$m_0(R)$ was taken from results of numerical simulations.
Dash-dotted lines show the second term in \eqnref{mp_generalform} with $c_p$ fitted.
Parameters: $\ku=1$ and $\st=0.1$ ({\bf a}), $\st=0.2$ ({\bf b}) and $\st=1$ ({\bf c}). Corresponding values of the phase-space correlation dimension $D_2=\{1.79,1.46,1.68\}$.
Panels {\bf c} and {\bf d}: Shows the moments $m_1(\R)$ ({\bf c}) and $m_2(\R)$ ({\bf d}) for $\st=0.2$. The scaling of asymptotic contributions is also shown.
Caustic contribution (dash-dotted), smooth contribution (dashed) and contribution at large separations (dotted).
}
\end{figure}

\Eqnref{mp_generalform} was derived in Ref.~\cite{Gus11b} in the white-noise limit.
Here we have shown how the same expression is obtained from matching
asymptotic forms of the joint distribution of separations and relative
velocities at finite Kubo numbers.
In Figs.~\figref{moments_1d_Ku1} and~\figref{moments_2d_Ku1} the theoretical result \Eqnref{mp_generalform} is compared to results
of numerical simulations at $\ku=1$ in one and two spatial dimensions.
We observe good agreement.

We conclude this subsection with three remarks. First, for small enough values of $\R$, we have $m_p(\R)\sim \R^{\min(p+D_2-1,d-1)}$.
For $p=0$ this shows that the spatial correlation dimension $d_2$ is bounded by $\min(D_2,d)$. This follows from the definition of the spatial correlation dimension, $m_0(\R\ll 1)\sim\R^{d_2-1}$.
We thus conclude: when the phase-space correlation dimension is less than the spatial dimension, it must coincide
with the spatial correlation dimension. This fact was observed in numerical simulations
of inertial particles suspended in incompressible random velocity fields \cite{Bec08}, and discussed
by Gustavsson and Mehlig \cite{Gus11b} in the white-noise limit.

Second, we remark that \eqnref{rho_asym_generaldim} expressed in Cartesian coordinates is symmetric under the interchange of $\Delta\ve v/\zstar$ and $\Delta\ve x$.
This implies:
\begin{align}
\rho(\V\ll\zstar)\equiv\int_0^\infty{\rm d}\R\rho(\V,\R)\sim\tilde b_0\V^{D_2-1}+\tilde c_0\V^{d-1}\,.
\end{align}
This expression corresponds to \eqnref{mp_generalform} with $p=0$, but the roles
of  $\Delta\ve v$ and $\Delta\ve x$ are interchanged.
This implies that the fractal dimension of the $\Delta\ve v$-coordinate is identical to the fractal dimension
of the $\Delta\ve x$-coordinate: $\rho(\V\ll\zstar)\sim \V^{{\rm min}(D_2,d)-1}=\V^{d_2-1}$, where $d_2$  is
the spatial correlation dimension, $m_0(R\ll 1 ) \sim R^{d_2-1}$.

Third, $m_1(\R)$ is closely related to the collision rate between two particles: the average in-going radial velocity between two spherical particles 
of radius $a$ into their collision sphere of radius $\R=2a$ is
\begin{align}
{\cal R}\approx m_1(2a)/2\,.
\eqnlab{collrate_approx}
\end{align}
The factor $1/2$ in \eqnref{collrate_approx} results from the fact that only in-going velocities ($\Vr<0$) contribute to the collision rate.
In writing \eqnref{collrate_approx} two approximations are made. First, the comparatively small asymmetry in the distribution of $\Vr$ for positive and negative values of $\Vr$ is neglected (here and throughout \Secref{3}). Second, re-collisions between the two particles contribute to \Eqnref{collrate_approx}, which may result in a large over counting 
in the collision rate for small values of $\st$~\cite{And07,Gus08}.
For small enough separations, the caustic $\R^{d-1}$-contribution in \eqnref{mp_generalform} becomes dominant if $p>d-D_2$.
In incompressible random and turbulent flows, $D_2>d-1$ and thus for small $\R$ the caustic contribution dominates $m_1(\R)$ and hence the collision rate for small particles.

\subsection{Smooth contribution to the moments}
\sseclab{moments_smooth}
The smooth part $m_p^{\mbox{\tiny (s)}}\sim b_p\R^{p+D_2-1}$  in Eq.~(\ref{eq:mp_generalform})
is due to the dynamics of particle pairs with small relative velocities in region $1$.
In general, this contribution is system-dependent and complicated: non-ergodic effects \cite{Gus11,Gus13}
may cause particle trajectories, separations and relative velocities to correlate with each other and with structures in the flow.
However, for incompressible random velocity fields, it turns out that non-ergodic effects in the dynamics of $\Delta\ve v$ are weak at small and intermediate values of $\ku$.
An ergodic treatment that neglect non-ergodic effects allows us to estimate $m_p^{\mbox{\tiny (s)}}$ to lowest order in $\ku$ as follows.
In the ergodic limit, we approximate $\ve u(\ve x_t,t) \approx \ve u(\ve x_0,t)$.
To lowest order in $\ku$ we approximate $\hat{\ve e}_R(t)\approx\hat{\ve e}_R(0)$, 
where $\hat{\ve e}_R(0)\equiv\Delta\ve x_0/\R_0$,  (see \Ssecref{DistCollVel}).
The dynamics of $\Vr$ to lowest order in $\ku$ follows from the linearised equation of motion (\ref{eq:deutsch_sep})
\begin{align}
\dotVr=\Delta\ve u(\ve x_0,\Delta\ve x_0,t)\cdot\hat{\ve e}_R(0)-\Vr\,.
\end{align}
Solving this equation yields:
\begin{align}
\Vr=\frac{1}{\st}\int_0^t{\rm d}t_1 \e^{(t_1-t)/\st}\Delta\ve u(\ve x_0,\Delta\ve x_0,t_1)\cdot\hat{\ve e}_R(t)\,,
\eqnlab{Vr_ergodic_solution}
\end{align}
where we have neglected the term containing the initial condition of $\Vr$ as we are interested in steady-state fluctuations
of $V_R$. Using this expression we calculate the moments $\overline{V_\R^p}$
conditional on $R_0 = |\Delta\ve x_0|$, to lowest order in $\ku$
(the conditional moments, also referred to as \lq particle-velocity structure functions\rq{} are 
different from the moments defined in Eq.~(\ref{eq:momentsmp}) and are discussed further in \Ssecref{structure_functions}).
From these moments we find that the distribution of $\Vr$ conditional on $R_0$ is a Gaussian, with variance (dropping the subscript in $R_0$)
\begin{align}
\overline{V_\R^2}=
\frac{2\ku^2\st^2}{d(1+\st)}\left(1+\frac{1}{\R}C'(\R,0)\right)
\eqnlab{VrSqr}
\end{align}
for an incompressible flow.
In \Eqnref{VrSqr}, $C(\R,t)$ is the correlation function~\eqnref{model_corrfun}, primes denotes derivatives with respect to $\R$,
and the result $C''(0,0)=-1$ was used (this follows from the normalisation adopted in \Secref{model}).
Using this Gaussian distribution we calculate the moments of absolute values of $\Vr$ conditional on $R$
\begin{align}
\overline{|\Vr|^p}&=\frac{2^{p/2}}{\sqrt{\pi}}\Gamma\left(\frac{p+1}{2}\right)\overline{V_\R^2}^{p/2}
\eqnlab{Vr_moments}
\end{align}
to lowest order in $\ku$.
Finally, we note that these conditional moments are related to the smooth contribution to the moments of $\Vr$ by
\begin{align}
m_p^{\mbox{\tiny (s)}}(\R)=m_0(\R)\overline{|\Vr|^p}\,.
\eqnlab{mp_smooth}
\end{align}
In particular, for small values of $\R$ (and $D_2<d$) we find
\begin{align}
m_p^{\mbox{\tiny (s)}}(\R\ll 1)\sim \R^{p+D_2-1}\underbrace{b_0\ku^p\st^p\frac{2^{p/2}}{\sqrt{\pi}}\Gamma\left(\frac{p+1}{2}\right)
\left[\frac{C''''(0,0)}{3d(1+\st)}\right]^{p/2}}_{b_p}\,,
\eqnlab{mp_smooth_smallR}
\end{align}
where we have expanded $C(\R,0)\sim 1-\R^2/2+C''''(0,0)\R^4/4!$ (valid for $R \ll 1)$.
\Eqnref{mp_smooth_smallR} relates the coefficients $b_p$ in (\ref{eq:mp_generalform}) with $p>0$ to $b_0$,
just as  \Eqnref{mp_smooth} relates $m_p^{\mbox{\tiny (s)}}(\R)$ to $m_0(\R)$.
We use \Eqnref{mp_smooth} to estimate the smooth contribution of the moments in \Eqnref{mp_generalform} and compare the result to numerical data in \Figref{moments_2d_Ku1}.
We observe good agreement.

In a single-scale flow with $D_2<d$, the 
probability density of separations $m_0(\R)$ can be approximated by its small $\R$ asymptote, $b_0\R^{D_2-1}$, for $\R<\Rstar$ and by its large $\R$ asymptote, $\R^{d-1}$, for $\R>\Rstar$. Here $\Rstar$
is the length scale at which the two asymptotes for very small and very large values of $\R$ in \eqnref{VrSqr} meet. For the two-dimensional incompressible model in \Secref{model}, $\Rstar=\sqrt{2}$.
Normalisation implies:
\begin{align}
b_0\approx d\Rstar^{d-D_2}(2/L)^d\,,
\eqnlab{b0}
\end{align}
where $L\gg 1$ is the system size.

\subsection{Singular contribution to the moments}
\sseclab{moments_singular}

The term $c_p\R^{d-1}$ in \Eqnref{mp_generalform} results from the formation of caustics.
As mentioned in \Secref{model}, caustic formation is an activated process, the formation rate ${\cal J}$ of caustics is of the form ${\cal J}\sim C(\st)\e^{-S(\mbox{\scriptsize$\st$})}$ for small values of $\st$.
In both white-noise flows and flows with finite values of $\ku$~\cite{Gus13} the \lq action' $S(\st)$ approaches infinity as $\st\to 0$ .
When $\st=0$ no caustics occur, ${\cal J}=0$. Formation of caustics is \lq activated' when ${\cal J}$ grows exponentially fast as $\st$ is increased.
It is expected that this activated behaviour is visible also in the coefficients $c_p$.
For small values of $\st$ we expect $c_p$ to approach zero exponentially fast.

The two asymptotes in \Eqnref{mp_generalform} are equal at the scale $\R_{\rm c}$:
\begin{align}
\R_{\rm c}=\left(\frac{c_p}{b_p}\right)^{1/(p+D_2-d)}\,.
\eqnlab{Rc}
\end{align}
This scale determines whether the smooth or singular contribution dominates in \Eqnref{mp_generalform}.
Due to the activated form of the caustic formation rate, $\R_{\rm c}$ approaches zero exponentially fast as $\st\to 0$, while $b_p$ approaches a non-zero value as $\st\to 0$.

The coefficient $c_p$ depends upon the Stokes number in a system-specific form that is determined by the 
degree of clustering and by how the tails of the distribution of radial velocities $\Vr$ are cut off for large values of $|\Vr|$ and small separations $\R$.
The tails of the distribution of $\Vr$ at small separations, $\rho(|\Vr|\gg\Vc,\R\ll 1)$, result from particle pairs in region $2$ originating at large separations, $\R_0\gg 1$. These particles are projected towards each other with approximately constant relative velocity \eqnref{region2_approx}.
When $\R\ll\R_0$ the distribution is approximately independent of $\Delta\ve x$ (apart from the geometrical prefactor $\R^{d-1}$) and the dependence on $\Delta\ve v$ is obtained from $\rho(|\Vr|\gg\Vc,\R_0)$.
Here $\R_0$ is taken along the \lq matching curve' described in \Secref{3}. In \Ssecref{DistCollVel} the power-law body of the distribution, $\rho(|\Vr|\ll\Vc,R\ll 1)$, was studied which resulted in the matching curve $\R_0\approx |\Vr|/\zstarx{\R}$ for $\R_0\ll 1$.
For the larger values of $\R_0$ studied in this subsection, this curve may be different.

In region $1$ we approximate the distribution $\rho(|\Vr|\ll\R_0,\R_0\ll 1)$ by a Gaussian in $\Vr$ with variance \eqnref{VrSqr} as in \Ssecref{moments_smooth}.
In the inertial range of a turbulent flow, Kolmogorov scaling
~\cite{Fri97} yields that the  prefactor $[1+C'(\R_0,0)/\R_0]$ in
\eqnref{VrSqr} scales as $\R_0^{2/3}$. 
%
% p.24 in the Lic then... ((3.16) with $C''(0,0)=-1$)
%
The matching curve can be estimated from a \lq variable-range projection' technique~\cite{Gus08b}, yielding 
the matching curve $\R_0\propto|\Vr|$ in the inertial range. The contribution to the distribution
from particles entering region $2$ in the inertial range thus becomes:
\begin{align}
\rho(|\Vr|\gg\Vc,\R\ll 1)\sim \R^{d-1}{\cal A}(\Vr)\exp\left[-{\cal C}V_\R^{4/3}\right]\,.
\eqnlab{rhoVr_inertial_range}
\end{align}
Here ${\cal A}(\Vr)$ is an unknown algebraic prefactor.
In the white-noise limit in one spatial dimension, ${\cal A}(\Vr)\sim V_\R^{-1}$~\cite{Gus08b}.
The result \eqnref{rhoVr_inertial_range} was first derived by Gustavsson {\em et al.}~\cite{Gus08b}, see also Ref.~\cite{Pan13}.
Dimensional analysis shows that the parameter ${\cal C}$ in \eqnref{rhoVr_inertial_range} is of the order $\sim(\gamma/\varepsilon)^{2/3}$~\cite{Gus08b} in dimensional units, 
where $\varepsilon$ is the dissipation rate per unit mass.

As $\st$ increases, more trajectories originating in the inertial range are projected to small separations.
If $\st$ is large enough, the contribution from the inertial range will dominate the values of the moments $m_p$.
In this case we approximate the distribution by extending \eqnref{rhoVr_inertial_range} to  the full range of $\Vr$.
Calculation of the moments using \eqnref{rhoVr_inertial_range}
and ignoring the prefactor ${\cal A}(\Vr)$  
yields the coefficient $c_p$ of the singular contribution
(quoted here in dimensional units for convenience):
\begin{align}
\frac{c_p}{c_0}\sim \Big(\frac{\eta}{\tau}\Big)^{p}\st^{p/2}\,.
\eqnlab{mp_inertial_range}
\end{align}
This result also follows from a dimensional argument~\cite{Vol80,Meh07}
(note that moments $\overline {|V_R|^p}$ of relative velocities
conditional on $R$ are given by  $c_p/c_0$ as $R\rightarrow 0$).
We remark that there is an unknown $\st$-dependence in $c_0$ in \Eqnref{mp_inertial_range}.
This implies that the $\st$-scaling of the collision rate \eqnref{collrate_approx} at large Stokes numbers
may differ from $\st^{1/2}$.

Finally, if $\R$ is much larger than the largest length scale in the flow ($\eta$ in single-scale flows 
or the upper cutoff $\Lambda$ of the inertial range in multi-scale flows), the matching curve is independent of $\R$.
It follows that the matching can be performed using \Eqnref{VrSqr} in the limit of $\R\to\infty$.
This implies that the distribution at small separations and for sufficiently large
values of $\Vr$, $\rho(|\Vr|\gg\Vc,\R\ll 1)$, is approximately:
\begin{align}
\rho(|\Vr|\gg\Vc,\R\ll 1)\sim\R^{d-1}\e^{-V_\R^2/(2\sigma^2)}\,,\hspace{0.5cm}\sigma^2=\frac{2\ku^2\st^2}{d(1+\st)}\,.
\eqnlab{single_scale_tail}
\end{align}
Using the same argument as for the inertial range above, we expect that \eqnref{single_scale_tail} dominates the contribution to the moments $m_p$ if $\st$ is large enough so that the distribution is dominated by caustics originating from separations much larger than the largest length scale of the system.
We find (the result is quoted in dimensional units):
\begin{align}
\frac{c_p}{c_0}\sim \Big(\frac{\eta}{\tau}\Big)^p
\,\st^{-p/2}\,.
\eqnlab{mp_largest_scale}
\end{align}
The $\st$-dependence in \eqnref{mp_largest_scale} is different from the $\st$-dependence that relates $c_p$ and $c_0$ when an inertial range is important \eqnref{mp_inertial_range}.
To estimate $c_0$, we use that the distribution of $\R$ is approximately uniform at all length scales, $m_0(\R)\sim\R^{d-1}$ (at small $\R$ because caustics are dominant there and at large $\R$ because then the variance \eqnref{VrSqr} is independent of $\Delta\ve x$).
This implies that $m_0(\R)\sim c_0\R^{d-1}$ and hence $c_0$ is approximately independent of $\st$ in \eqnref{mp_largest_scale}. In conclusion $c_p\sim \st^{-p/2}$ for large values of $\st$.

In contrast, for the case of caustics originating mainly from the inertial range we have a uniform distribution $m_0(\R)\sim\R^{d-1}$ for small enough values of $\R$, $\R\ll\R_{\rm e}$.
Here $\R_{\rm e}$ is the typical size of eddies such that its dimensional turnover time is comparable to $\gamma$. Dimensional analysis gives $\R_{\rm e}\sim\st^{3/2}$.
For $\R\gg\R_{\rm e}$, particles are advected, the distribution of $\Vr$ assumes the form of the distribution of $\Delta\ve u\cdot\hat{\ve e}_\R$. This distribution depends on $\R$ (as mentioned above the variance of $\Delta\ve u\cdot\hat{\ve e}_\R$ scales as $\R^{2/3}$ in the inertial range).
Hence, we do not expect the probability density of separations to be uniform for $\R_{\rm e}\ll\R\ll\Lambda$.
Because $\R_{\rm e}$ depends on the Stokes number, we expect the normalisation of $m_0(\R)$ and hence $c_0$ to depend on $\st$.

\subsection{Particle-velocity structure functions}
\sseclab{structure_functions}
In fluid dynamics \cite{Fri97} it is common to characterise the statistical properties of the flow field in terms of so-called \lq structure functions\rq{}.
Corresponding particle-velocity structure functions $S_p$ were analysed by Bec {\em et al.} \cite{Bec10}, see also \cite{Cen11,Sal12}. In terms of the moments $m_p$ we have
\begin{equation}
S_p(R)\equiv m_p(\R)/m_0(\R)  =\overline{|\Vr|^p} 
\end{equation}
%
% An alternative would need to define $S_p$ o
% when $\overline{|\Vr|^p}$ is defined and only use $S_p$
%
(we note that the structure functions $S_p$ are identical to the conditional moments $\overline{|\Vr|^p} $ discussed in
the two preceding subsections).
Using \Eqnref{mp_generalform} we find the following asymptotic behaviours for small values of $\R$ for the structure functions
\begin{equation}
S_p(\R)
\sim\left\{
\begin{array}{ll}
c_p/c_0 & \mbox{ if }D_2\ge d \mbox{ and }p\ge d-D_2\cr
\R^{p+D_2-d}b_p/c_0 & \mbox{ if }D_2\ge d \mbox{ and }p<d-D_2\cr
\R^{d-D_2}c_p/b_0 & \mbox{ if }D_2<d \mbox{ and }p\ge d-D_2\cr
\R^{p}b_p/b_0 & \mbox{ if }D_2<d \mbox{ and }p< d-D_2
\end{array}
\right.\,.
\eqnlab{Sp_generalform}
\end{equation}
When $\st>\st_{\rm c}$, where $\st_c$ is the critical value of $\st$ where $D_2$ approaches the spatial dimension from below, $D_2=d$, 
caustics give a constant contribution to the structure functions
[first case in \eqnref{Sp_generalform}].
We emphasise that caustics are important also for $\st<\st_{\rm c}$ [case $3$ in \eqnref{Sp_generalform}]~\cite{Gus12}.

As noted by Gustavsson and Mehlig~\cite{Gus11b} the results in \eqnref{Sp_generalform} qualitatively explain the form of the
structure functions of relative velocities at small separations observed in direct numerical simulations of particles in turbulent flows \cite{Bec10,Cen11} (see also \cite{Sal12}).
For $D_2<d$, the scaling exponents $\xi_p$ (defined from $S_p\sim\R^{\xi_p}$ as $\R\to 0$) in \eqnref{Sp_generalform} are $\xi_p=p$ if $p<d-D_2$ and $\xi_p=d-D_2$ if $p\ge d-D_2$.
This is what is seen in Fig.~4 in \cite{Bec10}: for a given value of $\st$, $\xi_p=p$ for small $p$ and saturates at a value $\xi_\infty$ which according to \eqnref{Sp_generalform} must be $\xi_\infty=d-D_2$. The last result is qualitatively consistent with the data for $D_2$ in \cite{Bec10}.
For $D_2\ge d$ (and $p>0$) \Eqnref{Sp_generalform} gives $\xi_p=0$ which is consistent with the numerical results in \cite{Bec10}.

Bec {\em et al.}~\cite{Cen11} compared the result \eqnref{Sp_generalform} to results of direct numerical simulations of turbulence.
Good agreement was obtained, except in the limit of small $\st$. This is expected because when $\st$ is small, caustics are rare and the length scales $\R<\R_{\rm c}$ are difficult to resolve in direct numerical simulations because $R_{\rm c}$ in \eqnref{Rc} approaches zero exponentially fast as $\st$ tends to zero.

For large enough values of $p$, $p\ge d-D_2$, and sufficiently small separations $\R$, $\R<\R_{\rm c}$, the caustic contribution dominates \Eqnref{Sp_generalform} and the scaling exponents become independent of $p$
\begin{align}
\xi_{p\ge d-D_2}\sim
d-d_2\,,
\eqnlab{xip_largep}
\end{align}
where $d_2=\min(D_2,d)$.
As mentioned in \Ssecref{moments}, direct numerical simulations of turbulent flows show that $D_2>d-1$ in incompressible turbulent flows~\cite{Bec07}.
This implies that $\eqnref{xip_largep}$ applies to the structure function with $p=1,2,\dots$.
Comparison to numerical data shows that \Eqnref{xip_largep} works approximately
for not too small values of $\st$ for $p=1$ (Fig.~3 in \cite{Bec10})
and for $p=2$ (Fig.~2 in \cite{Sal12}).

To conclude this section we comment on a common formulation in which the collision rate for particles
of radius $a$, \Eqnref{collrate_approx}, is rewritten in terms of the structure function $S_1$
\begin{align}
{\cal R}\approx 2\,a^{d-1}g(2a)S_1(2a)\,,
\end{align}
where $g(\R)\propto m_0(\R)/\R^{d-1}$ is the radial distribution function.
Further, since $g(\R)\sim\R^{d_2-d}$ it is often argued that clustering makes a substantial contribution to the collision rate.
But, as seen from \Eqnref{xip_largep}, also the structure function $S_1$ contains the factor $\R^{d-d_2}$. This factor cancels the power law
from $g(\R)$ for small values of $\R$.
We emphasise that it is $m_1(\R)$ that determines the collision rate, ${\cal R}\approx m_1(2a)/2$, \Eqnref{collrate_approx}. The singular contribution $c_1\R^{d-1}$ in \eqnref{mp_generalform} dominates $m_1(\R)$ for small values of $\R$ (or particle radius $a$).

\section{White-noise limit}
\label{sec:4}
In this section we show that the power laws of the distribution
of relative velocities at small separations \eqnref{rho_asym_generaldim}
are consistent with the solution of the Fokker-Planck equation describing
the corresponding distribution in the white-noise limit.
In the limit of $\ku\rightarrow 0$ and $\st\rightarrow \infty$ (so that $\epsilon^2\propto\ku^2\st$, \Eqnref{epsilon_def}, remains constant),
the particles experience the velocity field as a white-noise signal, and sample it in an ergodic fashion: the fluctuations of $\ve u(\ve x_t,t)$
(and its derivatives) along a particle trajectory $\ve x_t$ are indistinguishable from the fluctuations of $\ve u(\ve x_0,t)$ at the fixed position $\ve x_0$.
In this limit, the joint density of relative velocities $\Delta\ve v$ and separations $\Delta\ve x$ obeys
a Fokker-Planck equation \cite{vKa81}
with a $\Delta\ve x$-dependent diffusion matrix:
\begin{align}
\partial_t\rho
&=\sum_{i=1}^d\left[-\partial_{\Delta x_i}
(\Delta v_i\,\rho)+\partial_{\Delta v_i}(\Delta v_i\,\rho)\right]+\sum_{i,j=1}^d{\mathcal D}_{ij}(\Delta\ve x)\partial_{\Delta v_i}\partial_{\Delta v_j}\rho\,,\nn\\
{\mathcal D}_{ij}(\Delta\ve x) &\equiv \frac{1}{2}
\int_{-\infty}^\infty\!\!\!{\rm d}t \langle\Delta u_i(\ve x_0,\Delta\ve x,t)
\Delta u_j(\ve x_0,\Delta\ve x,0)\rangle\,.
\label{eq:FP}
\end{align}
Here the white-noise limit allows us to approximate $\Delta\ve u(\ve x,\Delta\ve x,t) \approx \Delta\ve u(\ve x_0,\Delta\ve x,t)=\ve u(\ve x_0+\Delta\ve x,t)-u(\ve x_0,t)$.
We adopt the same spherical coordinates as in \Ssecref{DistCollVel} in \eqnref{FP}: $\R,\phi_1,\dots,\phi_{d-1}$ with unit vectors $\hat{\ve e}_R,\hat{\ve e}_{\phi_1},\dots,\hat{\ve e}_{\phi_{d-1}}$.
As in \Ssecref{DistCollVel} we project $\ve z\equiv\Delta\ve v/\R$ onto the unit vectors of the spherical coordinate system, $z_\alpha\equiv\Delta\ve v\cdot\hat{\ve e}_\alpha/\R$ with $\alpha=R,\phi_1,\dots,\phi_{d-1}$.
We change coordinates from $\Delta\ve v$ to $z_\alpha$ in \eqnref{FP}.
Further, we consider the limit $\R\to 0$ and expand $\Delta\ve u$ for small separations in terms of the flow-gradient matrix $\ma A$:
\begin{align}
\Delta\ve u(\ve x_0,\Delta\ve x,t)=\ma A(\ve x_0,t)\Delta\ve x\,.
\eqnlab{Deltau_smallR}
\end{align}
In spherical coordinates the diffusion matrix in \eqnref{FP}
\begin{align}
{\cal D}_{\alpha\beta}&\equiv
\frac{\R^2}{2}\int_{-\infty}^\infty\!\!\!{\rm d}t \langle(\hat{\ve e}_\alpha\transpose\ma A(\ve x_0,t)\hat{\ve e}_R)(\hat{\ve e}_\beta\transpose\ma A(\ve x_0,0)\hat{\ve e}_R)\rangle
\end{align}
is diagonal for isotropic flows and has non-zero components ${\cal D}_{RR}=\epsilon^2R^2$ and ${\cal D}_{\phi_i\phi_i}=\epsilon^2\Gamma R^2$.
Here $\Gamma$ is a parameter characterising the degree of compressibility of flows in two or three spatial dimensions (see Table $1$ in \cite{Gus13b}). It equals $\Gamma=(d+1)/(d-1)$ for incompressible flows and $\Gamma=1/3$ for potential flows.
The radial diffusion constant $\epsilon^2$ equals $\epsilon^2=\ku^2\st(d+2)/(d(1+(d-1)\Gamma))$ for the model in \Secref{model} \cite{Gus08}.
The values of $e_d$ quoted below \Eqnref{epsilon_def} follow from this expression.

Assuming boundary conditions consistent with spherical symmetry, the distribution $\rho$ becomes independent of the angular variables that can thus 
be integrated away from \Eqnref{FP}. The corresponding steady-state Fokker-Planck equation for small values of $\R$ is:
\begin{align}
z_R\partial_R R\rho&=\partial_{z_R}(z_R+z_R^2-\sum_{i=1}^{d-1}z_{\phi_i}^2)\rho+\epsilon^2\partial_{z_R}^2\rho
\nn\\&
+\sum_{i=1}^{d-1}\left[(1+2z_R)\partial_{z_{\phi_i}}z_{\phi_i}\rho
+\epsilon^2\Gamma\partial_{z_{\phi_i}}^2\rho\right]\,.
\eqnlab{FP_Rz_general_dim}
\end{align}
This equation is equivalent to those studied Wilkinson and Mehlig \cite{Wil03}, Mehlig and Wilkinson \cite{Meh04}, and Wilkinson {\em et al.} \cite{Wil07}
in one, two and three spatial dimensions respectively.

Inserting the ansatz
\begin{equation}
\rho_\mu=g_\mu(\R) Z_\mu(z_R,z_{\phi_1},\dots,z_{\phi_{d-1}})
\eqnlab{rho_sep_ansatz}
\end{equation}
with separation constant $\mu$ into \Eqnref{FP_Rz_general_dim}, we obtain in the limit of $\R\to 0$:
\begin{align}
g_\mu(\R)&=\R^{\mu-1}\,,\\
\mu z_RZ_\mu &\!=\! \big[\partial_{z_R}(z_R\!+\!z_R^2\!-\!\sum_{i=1}^{d-1}z_{\phi_i}^2)\!+\!\epsilon^2\partial_{z_R}^2\!
+\!\sum_{i=1}^{d-1}\big((1\!+\!2z_R)\partial_{z_{\phi_i}}z_{\phi_i}\!+\!\epsilon^2\Gamma\partial_{z_{\phi_i}}^2\big)\big]Z_\mu\,.
\label{eq:general_z_FP}
\end{align}
For $\mu=0$, Eq.~(\ref{eq:general_z_FP}) determines
the rate at which singularities in the
particle-velocity
gradients $\partial\ve v/\partial\ve x$ are created \cite{Wil03,Wil05}.

In the following  we investigate the joint
distribution of finite differences in positions and velocities.
This distribution is determined by solutions of (\ref{eq:general_z_FP})
with finite values of $\mu$.
If the stochastic driving is neglected ($\epsilon=0$), \Eqnref{general_z_FP} has the solution
\begin{align}
Z_\mu=|z_R+\Z^2|^{\mu/2-d}\left(\frac{1+2z_R+\Z^2}{\Z^2}\right)^{\mu/4}\!\!\!\!f\Big(\frac{z_{\phi_2}}{z_{\phi_1}},\frac{z_{\phi_3}}{z_{\phi_1}},\dots,\frac{z_{\phi_{d-1}}}{z_{\phi_1}},\frac{z_R+\Z^2}{z_{\phi_1}}\Big)
\eqnlab{nonoise_FPsolution}
\end{align}
where $\Z^2=z_R^2+z_{\phi_1}^2+\dots+z_{\phi_{d-1}}^2$.
As argued in \Secref{3}, the distribution of $\Delta\ve v$ and $\Delta\ve x$ is independent of the direction of $\Delta\ve v$ in the limit of $\V\gg\R$.
We thus expect that in the limit $\Z=\V/\R\to\infty$ the function $Z_\mu$ depends upon $\Z$ only.
This implies $f=\const$ in \eqnref{nonoise_FPsolution} and we find
\begin{align}
Z_\mu\sim\Z^{\mu-2d}
\eqnlab{nonoise_FPsolution_largeZ}
\end{align}
for large values of $\Z$.
Large values of $\Z$ correspond to the dynamics in region 2 in Figs.~\figref{trajek}--\figref{regions_universal}.

For the distribution \eqnref{rho_sep_ansatz} to be integrable at $\R=0$ we must require $\mu>0$.
We interpret \Eqnref{general_z_FP} as an eigenvalue problem, with eigenvalues $\mu$ and right eigenfunctions $Z_\mu$.
We use a \lq weight function' $\omega=z_R$ and the corresponding left eigenfunctions $\Zleft{\mu}$ are on the form $\Zleft{\mu}\sim \Z^{-\mu}$ for large values of $\Z$. Here we have assumed that $\Zleft{\mu}$ depends on $\Z$ only.
A complete set of allowed solutions $Z_\mu$ is obtained by requiring integrability of $\omega Z_\mu \Zleft{\nu}$ at large values of $\Z$ for all allowed values of $\mu$ and $\nu$.
We find that the allowed eigenvalues $\mu$ must satisfy $\muc\le \mu\le \muc+d-1$ where $\muc$ is a yet undetermined lowest eigenvalue.
In general one expects that the distribution $\rho(\Delta\ve v,\R)$ is obtained by summing or integrating the solution \eqnref{rho_sep_ansatz} over a discrete or continuous subset of allowed eigenvalues.
In one spatial dimension there is only one allowed eigenvalue as we show in \Ssecref{FP1d_etainf}.
Results of numerical simulations of our model show that this
is true in higher dimensions too, for small values of $\R$.
Changing back to the variables $\Delta\ve v=\ve z\R$ and $\R$ yields in this case:
\begin{align}
\rho(\Delta\ve v,\R)
&\sim\rho_\muc(z_R,z_{\phi_1},\dots,z_{\phi_{d-1}},\R)\prod_\alpha\frac{\partial z_{\alpha}}{\partial v_\alpha}
=\R^{d-1}\V^{\muc-2d}\,.
\eqnlab{rhoFP2}
\end{align}
Comparing this power-law solution to the universal power-law form \eqnref{rho_asym_generaldim} in region $2$
allows us to conclude that $\muc=D_2$.

For $\Z\ll 1$ by contrast, the quadratic terms and the left hand side in \Eqnref{general_z_FP} may be neglected and the equation for $Z_\mu$ is that of 
an Ornstein-Uhlenbeck process with a Gaussian steady-state solution in $z_R$, $z_{\phi_1}$,\dots,$z_{\phi_{d-1}}$, with variances $\epsilon^2$ for $z_R$ and $\Gamma\epsilon^2$ for $z_{\phi_i}$ ($i=1,\dots,d-1$).
The solution to the Fokker-Planck equation in region 1 with $\mu=\muc=D_2$ becomes
\begin{align}
\rho(\Delta\ve v,\R)
&\sim\R^{D_2-d-1}\exp\left[-\frac{V_\R^2}{2\epsilon^2\R^2}-\sum_{i=1}^{d-1}\frac{V_{\phi_i}^2}{2\epsilon^2\Gamma\R^2}\right]\,.
\eqnlab{rhoFP1}
\end{align}
As $\V/\R\to 0$ this solution approaches $\R^{D_2-d-1}$ which is identical to the universal asymptotic solution in region 1 in \eqnref{rho_asym_generaldim_ansatz}.

In summary the asymptotic distribution \eqnref{rho_asym_generaldim} is consistent with the solutions of the Fokker-Planck equation \eqnref{FP} in different asymptotic limits.
The white-noise results \eqnref{rhoFP2} and \eqnref{rhoFP1} were used to derive the results in general spatial dimension in~\cite{Gus11b}.

\section{One-dimensional model}
\seclab{onedim}

In this section we show how to compute the joint distribution of relative velocities
and separations for particles suspended in a random velocity field in one spatial dimension \cite{Wil03}. We also show how to calculate the matching parameters $D_2$ and $\zstar$.
In \Ssecref{FP1d_etainf} we solve the one-dimensional white-noise problem analytically in the limit of $\R/\eta\to 0$ (in dimensional units), 
which makes it possible to calculate $D_2$ and $\zstar$.
In \Ssecref{FP1d_finite} we investigate how the distribution is modified in a system with finite correlation length $\eta$.

\subsection[Determining D2 and z* in the white-noise limit]{Determining $D_2$ and $\zstar$ in the white-noise limit}
\sseclab{FP1d_etainf}

The main result of \Secref{3}, \Eqnref{rho_asym_generaldim}, is an asymptotic approximation
of the form of the joint distribution function $\rho(\Delta v, \Delta x)$. It contains two as yet undetermined
parameters, the phase-space correlation dimension $D_2$ and the matching parameter $z^\ast$.
In this subsection we demonstrate how these two parameters
can be computed from first principles in the white-noise limit.
We note that a brief account of some of the results described in this subsection was published previously by Gustavsson and Mehlig in Ref.~\cite{Gus11b}.

The distribution $\rho(\Delta v, \Delta x)$ is given by the steady-state solution of
the Fokker-Planck equation \eqnref{FP}. We can compute the solution of this equation as a series expansion in small values of $\R=|\Delta x|$.
The lowest-order equation of this series expansion is solved in this subsection.
The corresponding solution is valid in the limit of infinite correlation length, $\eta\to\infty$, 
or equivalently as $\R\to 0$ (this is a consequence of the dimensionless units adopted in \Secref{model}).
In this limit we can calculate the parameters $D_2$ and $\zstar$.
However, because it is assumed that $\eta$ tends to infinity, the tails of the distribution are not cut-off, the power laws in $\Delta v$ and $\Delta x$ extend to infinity.
In \Ssecref{FP1d_finite} we show how to resolve this problem by solving the equations to higher orders in $\R$. This makes it possible to calculate the power-law cut offs for large $\R$ and large $\V$.

\begin{figure}
\includegraphics[width=6.5cm,clip]{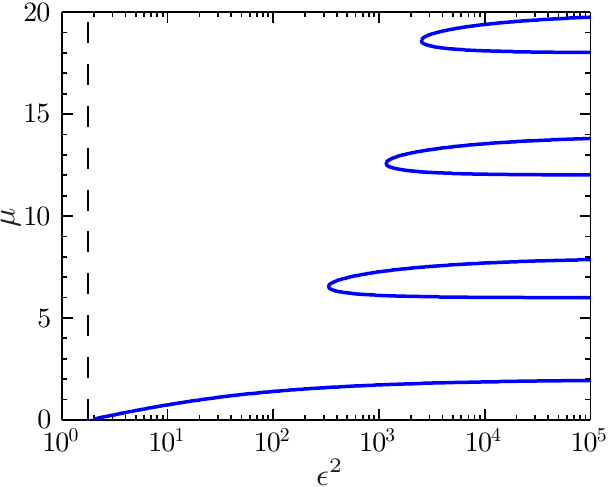}
\caption{\label{fig:muc}
Positive eigenvalues of \Eqnref{Zmu}  allowed by the condition \eqnref{A} shown as a function
of $\epsilon^2$ ($\epsilon>1.33$ so that the maximal Lyapunov exponent is positive).
}
\end{figure}

In one spatial dimension, separation of variables in
\eqnref{FP_Rz_general_dim} takes the form:
\begin{align}
g_\mu(\R)=\R^{\mu-1}
\,,\hspace{0.5cm}
\mu z Z_\mu = \partial_z(z+z^2+\epsilon^2\partial_z)Z_\mu(z)\,.
\label{eq:Zmu}
\end{align}
Certain values of $\mu$ and $\epsilon$ allow for exact solutions of \eqnref{Zmu}.
Expansion in $\mu$ around such a solution makes it possible to find
approximate
solutions of \eqnref{Zmu}.
A related expansion
was used by Schomerus {\em et al.} \cite{Sch02} to calculate moments of the  finite-time Lyapunov exponent for
particles accelerated in a random time-dependent potential.
One example of an exact solution of \eqnref{Zmu} is obtained for $\mu=0$:
\begin{align}
Z_0 &= A_0\,\e^{-V(z)}\int_{-\infty}^z \e^{V(z')}d\,z'
\end{align}
with $V(z)=\epsilon^{-2}\,(z^3/3+z^2/2)$ and the boundary condition $Z_\mu(z=-\infty)=0$
(see \cite{Wil03}).
For other values of $\mu$, a perturbative solution can be found by
expanding $Z_\mu(z)=Z(\mu,z)$ in powers of $\mu$:
\begin{align}
Z(\mu,z)=\sum_{n=0}^\infty\frac{1}{n!}\partial^n_\mu Z(\mu=0,z)\mu^n\,.
\label{eq:Zmu_serie}
\end{align}
Repeatedly differentiating the Fokker-Planck equation (\ref{eq:Zmu}) w.r.t. $\mu$ we find upon inserting $\mu=0$:
\begin{align}
0&=\left.\partial^n_\mu\left\{\partial_z(z+z^2+\epsilon^2\partial_z)Z(\mu,z)-\mu zZ(\mu,z)\right\}\right|_{\mu=0}\nn\\
&=\partial_z(z+z^2+\epsilon^2\partial_z)\partial^n_\mu Z(\mu=0,z)-n z\partial^{n-1}_\mu Z(\mu=0,z)\,.
\label{eq:Z_inhomogeneous}
\end{align}
To any order, this is an inhomogeneous version of the original Fokker-Planck equation (\ref{eq:Zmu}).
The solution to (\ref{eq:Z_inhomogeneous}) at order $n$ becomes (in terms of the solution at order $n-1$):
\begin{align}
\newcommand{\mzero}{\ma{\mathbb{O}}}
\newcommand{\tr}{\mathrm{Tr}}
\newcommand{\lep}{\left.}
\newcommand{\rep}{\right.}
\newcommand{\lp}{\left(}
\newcommand{\rp}{\right)}
\newcommand{\lap}{\left<}
\newcommand{\rap}{\right>}
\newcommand{\lbp}{\left\{}
\newcommand{\rbp}{\right\}}
\newcommand{\half}{\frac{1}{2}}
\newcommand{\Binom}[2]{{}_{#1}\mathcal{C}_{#2}}
\newcommand{\Ordo}{\mathcal{O}}
\partial^n_\mu Z(\mu=0,z)&=n\,\epsilon^{-2}\,\e^{-V(z)}\int_{-\infty}^{z}{\rm d}{z_1}\e^{V(z_1)}\int_{-\infty}^{z_1}{\rm d}{z_2}\,z_2\,\partial^{n-1}_\mu Z(\mu=0,z_2)\,,\mbox{ for }n>0\nn\\
\partial^n_\mu Z(\mu=0,z)&=A_0\,\e^{-V(z)}\int_{-\infty}^{z}{\rm d}{z_1}\,\e^{V(z_1)}\,,\mbox{ for }n=0\,.
\label{eq:Z_terms_recursive}
\end{align}
By making the variable transformations $z_i=\sum_{j=0}^i t_j$ (with $t_0=z$) in (\ref{eq:Z_terms_recursive}) and inserting
the result into (\ref{eq:Zmu_serie}) we obtain \cite{Gus11b}
\begin{align}
Z_\mu(z) &= \sum_{k=0}^\infty
\Big(\frac{\mu}{\epsilon^2}\Big)^k \int_{-\infty}^0
\!\!\!{\rm d}t_1\cdots {\rm d}t_{2k+1}
\Big(\prod_{i=1}^k\sum_{j=0}^{2i}t_j\Big)
\exp\Big(-\!\sum_{i=0}^{2k+1}(-1)^iV(\sum_{j=0}^it_j)\Big)\,.
\label{eq:Z2}
\end{align}
Eq.~(\ref{eq:Z2}) represents
the general solution to (\ref{eq:Zmu}) for a given value of $\mu$ and with initial
condition $Z_\mu(z=-\infty)=0$.
The appropriate choice of $\mu$ is determined by the boundary conditions.
In order to find the allowed values of $\mu$ we consider the large-$|z|$ asymptote of
$Z_\mu(z)$. It is most easily obtained by inspecting (\ref{eq:Zmu}). When $|z|$ is large,
then the first term on
the right hand side of the second equation (\ref{eq:Zmu}), $\partial_z (z Z_\mu(z))$, can be neglected and the resulting equation
is solved by Kummer functions with the asymptotic behaviour:
\begin{equation}
Z_\mu(z)\sim A^\pm_\mu\, (\pm z)^{\mu-2}\, \mbox{ for } z \rightarrow  \pm \infty\,,
\label{eq:asy}
\end{equation}
where $A^+_\mu$ and $A^-_\mu$ are constants.
Eq.~(\ref{eq:asy})  suggests that
$\rho(\Delta v,\R)$ approaches power-law form
as $\R\rightarrow 0$
\begin{equation}
\label{eq:xz}
\rho(\Delta v, \Delta x)
=\R^{\mu-2} Z_{\mu}({\Delta v}/{\Delta x})
\sim A^\pm_\mu\,(\pm \Delta v)^{\mu-2}
\end{equation}
for positive and negative values of $\Delta v$, respectively.
Now consider exchanging two particles in a pair:
$\Delta x \rightarrow -\Delta x$ and $\Delta v \rightarrow -\Delta v$.
The steady-state solution to (\ref{eq:FP}) is invariant under this
exchange. We must therefore require that
\begin{equation}
\label{eq:A}
A^+_\mu=A^-_\mu\,.
\end{equation}
This results in a condition on the allowed values of $\mu$.
Numerically evaluating (\ref{eq:Z2}) for different values of $\mu$,
and choosing $\mu$ such that condition (\ref{eq:A}) is satisfied
yields a discrete set of possibly allowed $\mu$-values.
These are shown in \Figref{muc}.
Note that the allowed
values of $\mu$ depend on the parameter $\epsilon$.
Note also that as $\epsilon\to\infty$ then $Z_\mu(z)$ approaches $U((2-\mu)/3,2/3,-z^3/(3\epsilon^2))$. Here $U(a,b,x)$ is the confluent hypergeometric
Kummer function of the second kind.
This solution satisfies the condition \eqnref{A} with positive $\mu$ if $\mu=2+6n$ or $\mu=6+6n$ with $n=0,1,2,\dots$.
Such staggered ladder spectra were obtained in Refs.~\cite{Arv06,Bez06} in a different context.
For the steady-state distribution $\rho_\mu(z,\R)\sim\R^{\mu-1}Z_\mu(z)$ to be integrable at $\R=0$
we must require that $\mu > 0$.

\begin{figure}
\includegraphics[width=13.5cm]{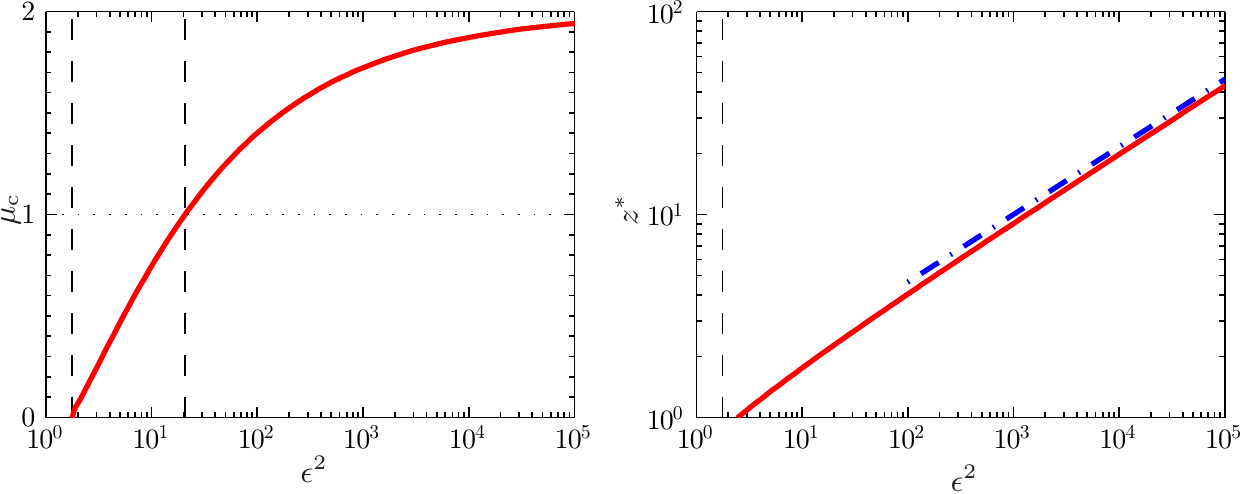}
\caption{\label{fig:data_mu_zstar}
Left: The allowed value of $\mu$, $\muc=D_2$, as a function of $\epsilon^2$,
obtained from (\ref{eq:Z2}) and the condition (\ref{eq:A}).
Right: the parameter $\zstar$ as a function of $\epsilon^2$,
obtained from (\ref{eq:zstar_cond}).
In both panels, the critical value of $\epsilon$ for which
the path-coalescence transition \cite{Wil03} occurs, $\epsilon^2\approx 1.77$,
is shown as a dashed line. In the left panel, the second critical value \cite{Gus11b} where $\mu_{\rm c}$ crosses unity, $\epsilon^2\approx 20.7$,
is also shown. Finally, in the right panel, the power law $\epsilon^{2/3}$ is indicated by a blue dashed-dotted line.
}
\end{figure}

In order to select a complete set of functions $Z_\mu$ we consider the left eigenfunctions $\Zleft{\mu}(z)$ of (\ref{eq:Zmu}) and weight $\omega=z$ as in \Secref{4}.
As noted above, the large-$z$ asymptotics of $Z_\mu$ is $Z_\mu(z)\sim A_\mu|z|^{\mu-2}$ and in a similar manner it is possible to show that $\Zleft{\mu}(z)\sim B^\pm_\mu |z|^{-\mu}$ for some constants $B^\pm_\mu$ as $|z|\to\pm\infty$. For values of $\mu$ such that $A^+_\mu=A^-_\mu$ it turns out that $B^+_\mu=B^-_\mu$ (up to the numerical precision of our solutions) and thus $\Zleft{}(z)$ is symmetric for large values of $|z|$. This gives
$\omega Z_\mu(z)\Zleft{\nu}(z)\sim z|z|^{\mu-\nu-2}$ which is integrable for large $z$ when $\mu=\nu$,
% (because of the antisymmetry for large $z$), 
but not when $\mu>\nu$.
This implies that there is only one allowed eigenfunction $Z_\muc$ with eigenvalue $\muc\equiv\muc(\epsilon)$.
% As in \Secref{4}, to find the correct value of $\muc$ we compare the solution $Z_\muc$ to the universal power law scalings in \Eqnref{rho_asym_generaldim} (these are derived in the limit of small $\R$ studied in this section).
Inserting $d=1$ into \eqnref{rho_asym_generaldim} and comparing to Eq.~(\ref{eq:xz}) allows us to identify $\muc=D_2$.
In the range of $0<\muc\le 2$ and for values of $\epsilon$ so that the maximal Lyapunov exponent is positive, only one solution $\muc$ exists.
This solution is shown as a function of $\epsilon^2$ in the left panel of Fig.~\ref{fig:data_mu_zstar}.
In conclusion,
\begin{align}
\rho(\Delta v,\Delta x)=\left.\frac{\rho_{D_2}(z,R)}{\R}\right|_{z=\Delta v/\Delta x,\,R=|\Delta x|}=|\Delta x|^{D_2-2}Z_{D_2}(\Delta v/\Delta x)
\eqnlab{rho1d_snallR_final}
\end{align}
is the exact form of the distribution of separations and relative velocities in the one-dimensional white-noise model in the limit of small $\R=|\Delta x|$.
Taking the limits of $\V\ll\R$ and $\V\gg\R$ in \Eqnref{rho1d_snallR_final} recovers the asymptotic power laws of region 1 and region 2 in \Eqnref{rho_asym_generaldim} with the exception that the tails are not cut off in \eqnref{rho1d_snallR_final}.
The solution \eqnref{rho1d_snallR_final} contains no cut off (the power laws extend to infinity) because we assumed that $\eta\to\infty$ in this subsection. 
An improved solution that resolves this problem is described in \Ssecref{FP1d_finite}.

Fig.~\ref{fig:distributions1d_boundary} shows results of numerical simulations of Eq.~(\ref{eq:deutsch}), compared with \Eqnref{rho1d_snallR_final}.
We see that \eqnref{rho1d_snallR_final} predicts the distribution correctly for small enough values of $\R$ and $\V$, but fails for large values of $V$.

We estimate the second parameter $\zstar$ occurring in \eqnref{rho_asym_generaldim} by comparing \eqnref{rho_asym_generaldim} with \Eqnref{rho1d_snallR_final} for small values of $\R$.
We obtain two conditions (resulting from comparison at small and at large values of $\Delta v/\Delta x$, respectively):
\begin{align}
\label{eq:eq1}
&{\cal N}\R^{D_2-2}=\R^{D_2-2}Z_{D_2}(0)\,,\\
\label{eq:eq2}
&{\cal N}(\V/\zstar)^{D_2-2}=A_{D_2}\V^{D_2-2}\,.
\end{align}
Here ${\cal N}$ is a global normalisation factor for \Eqnref{rho_asym_generaldim}.
The parameter $z^\ast$ is determined by the conditions (\ref{eq:eq1}) and (\ref{eq:eq2}).
We find
\begin{align}
\zstar=\left(\frac{Z_{D_2}(0)}{A_{D_2}}\right)^{1/(D_2-2)}
\label{eq:zstar_cond}
\end{align}
where the ratio $Z_{D_2}(0)/A_{D_2}$ is obtained from Eq.~(\ref{eq:Z2}).
The resulting values of $z^\ast$ are shown in the right panel of Fig.~\ref{fig:data_mu_zstar} as a function of $\epsilon^2$.

\subsection[Determining the large-V cutoff in the white-noise limit]{Determining the large-$\V$ cutoff in the white-noise limit}
\sseclab{FP1d_finite}

In \Ssecref{FP1d_etainf} we showed how to calculate the distribution of relative velocities and separations in a one-dimensional white-noise model in the limit of $\R=|\Delta x| \to 0$ ($\eta\to\infty$ in dimensional units).
In this subsection we show how this distribution is modified for
% a finite system size
finite values of $\eta$.
In this case the distribution at finite values of $\R$ can be calculated in terms of a series expansion in small $\R$. The solution \eqnref{rho1d_snallR_final} is the lowest-order solution of this expansion.
Consider one spatial dimension. We substitute
\begin{equation}
\zeta = \frac{\sign(\Delta x)\epsilon\Delta v}{\sqrt{{\cal D}(\R)}}
\eqnlab{zeta_def}
\end{equation}
with $\R>0$ into \Eqnref{FP}. The steady-state
form of the resulting equation for the distribution $\rho(\zeta,\R)$ is:
\begin{align}
\partial_{\R}\frac{\zeta\sqrt{{\cal D}(\R)}}{\epsilon}\rho&=\partial_{\zeta}\left(\zeta+\zeta^2\frac{\partial_{\R} {\cal D}(\R)}{2\epsilon\sqrt{{\cal D}(\R)}}\right)\rho
+\epsilon^2\partial_{\zeta}^2\rho\,.
\eqnlab{start_FP}
\end{align}
In \Ssecref{FP1d_etainf} we considered the limit of small separations $\R\to 0$, so that ${\cal D}(\R)\to\epsilon^2\R^2$ and $\zeta\to z=\Delta v/\Delta x$.
In this limit, after a separation of variables, \Eqnref{start_FP} simplifies to \Eqnref{Zmu}, and the solution is on the form $\rho(\zeta,\R)\to \R^{D_2-1}Z_{D_2}(\zeta)$, according to \eqnref{rho1d_snallR_final}.

For finite values of $\Delta x$ we expand
\begin{align}
\sqrt{{\cal D}(\R)}&=\epsilon\sum_{j=0}^\infty \beta_j\R^{2j+1}\,,
\eqnlab{sqrtDX_finiteX_expansion}
\end{align}
where $\beta_0=1$ from the definition of $\epsilon$.
Note that the series expansion of $\sqrt{{\cal D}(\R)}$ has a finite radius of convergence.
Depending on the form of ${\cal D}(\R)$ and the value of $\epsilon$, the expansion \eqnref{sqrtDX_finiteX_expansion} may fail at too large values of $\R$.
In the following we restrict our analysis to values of $\R$ such that the expansion \eqnref{sqrtDX_finiteX_expansion} converges.

Next we expand $\rho(\zeta,\R)$ for small values of $\R$
 as a sum of power-law terms with different powers shows that the smallest power must be $D_2-1$ in accordance with the $\R\to 0$ solution.
Higher expansion powers in $\R$ are chosen to match the powers of the expansion terms coming from substitution of \eqnref{sqrtDX_finiteX_expansion} into \Eqnref{start_FP}:
\begin{align}
\rho(\zeta,\R)&=\sum_{i=0}^\infty \alpha_i(\zeta)\R^{D_2-1+2i}\,,
\eqnlab{rho_finiteX_ansatz}
\end{align}
where $\alpha_i(\zeta)$ are functions to be determined.
We expand the Fokker-Planck equation \eqnref{start_FP} in $\R$ using \eqnref{sqrtDX_finiteX_expansion} and \eqnref{rho_finiteX_ansatz}.
Collecting terms of order $\R^{D_2-1+2k}$ yields a recursion of equations for $\alpha_k(\zeta)$
\begin{align}
&\left[-(D_2+2k)\zeta+\partial_{\zeta}(\zeta+\zeta^2)+\epsilon^2\partial_{\zeta}^2\right]\alpha_k\nn\\
&\hspace*{1cm}=\sum_{i=0}^{k-1}\beta_{k-i}\left[(D_2+2k)\zeta-(2k-2i+1)\partial_{\zeta}\zeta^2\right]\alpha_i\,.
\eqnlab{Alt_FP_smallR_recursion}
\end{align}
In summary, the solution $\rho(\zeta,R)$ of \eqnref{start_FP} is given by the ansatz \eqnref{rho_finiteX_ansatz} where $\alpha_k(\zeta)$ 
are determined recursively by the solution of \Eqnref{Alt_FP_smallR_recursion}.
In \Appref{ak_calculation} we show how to solve \Eqnref{Alt_FP_smallR_recursion} to obtain $\alpha_k(\zeta)$ .
Each solution $\alpha_k(\zeta)$ has a normalisation that must be determined. This normalisation is found by analysing the asymptotic behaviour of $\alpha_k(\zeta)$ for large values of $\zeta$. For large negative $\zeta$ we find $\alpha_k(\zeta)\sim\alpha_k^-|\zeta|^{D_2-2+2k}$, and $\alpha_k^-$ determines the normalisation of $\alpha_k(\zeta)$,
see \Appref{ak_calculation}.

\begin{figure}
\includegraphics[width=13.5cm,clip]{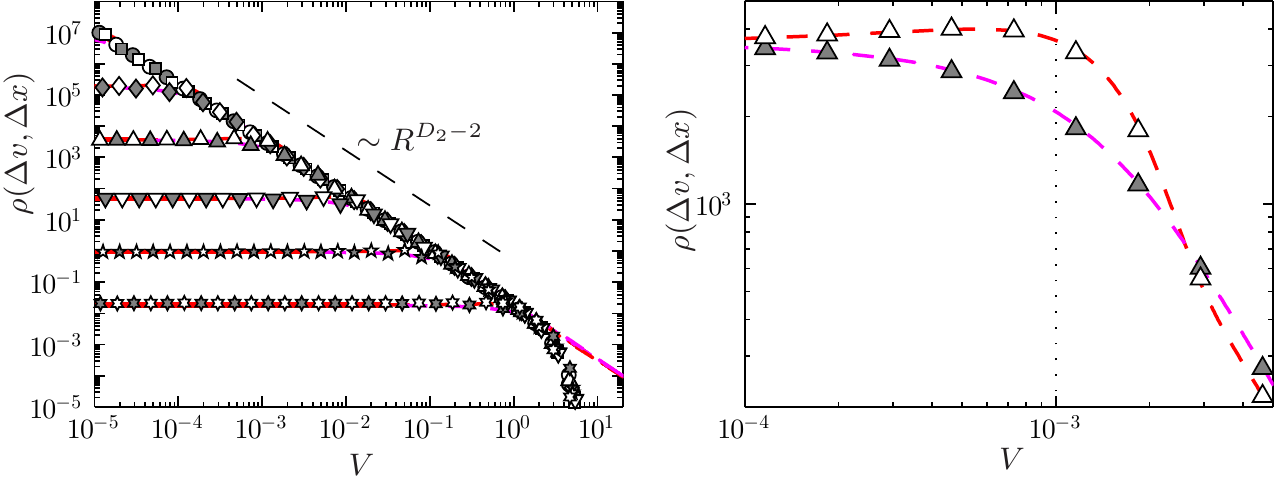}
\caption{\label{fig:distributions1d}
Left: Distribution $\rho(\Delta v,\Delta x)$ in one spatial dimension in the white-noise limit for different values of $\Delta x$: $\Delta x=10^{-6}$ ($\circ$), $10^{-5}$ ($\square$), $10^{-4}$ ($\diamond$), $10^{-3}$ ($\vartriangle$), $0.01$ ($\triangledown$), $0.1$ ($\star$) and $1$ ($\ast$).
Symbols show results of numerical simulations of the model described in \Secref{model}.
White/shaded symbols show data for positive/negative values of $\Delta v$.
Red/magenta dashed lines show the small $\Delta x$-theory \eqnref{rho1d_snallR_final} for positive/negative values of $\Delta v$ (norm fitted).
The values of $D_2\approx 0.24$ and $z^\ast\approx 1.09$ were taken from Fig.~\ref{fig:data_mu_zstar}.
Parameters: $\ku=0.1$ and $\st=100$ (white-noise limit) so that $\epsilon^2=3$.
Right: Zoom in of $\rho(\Delta v,\Delta x=0.001)$ around the matching scale $\V=\zstar\R$ (black dotted).
}
\vspace{0.5cm}
\includegraphics[width=13.5cm]{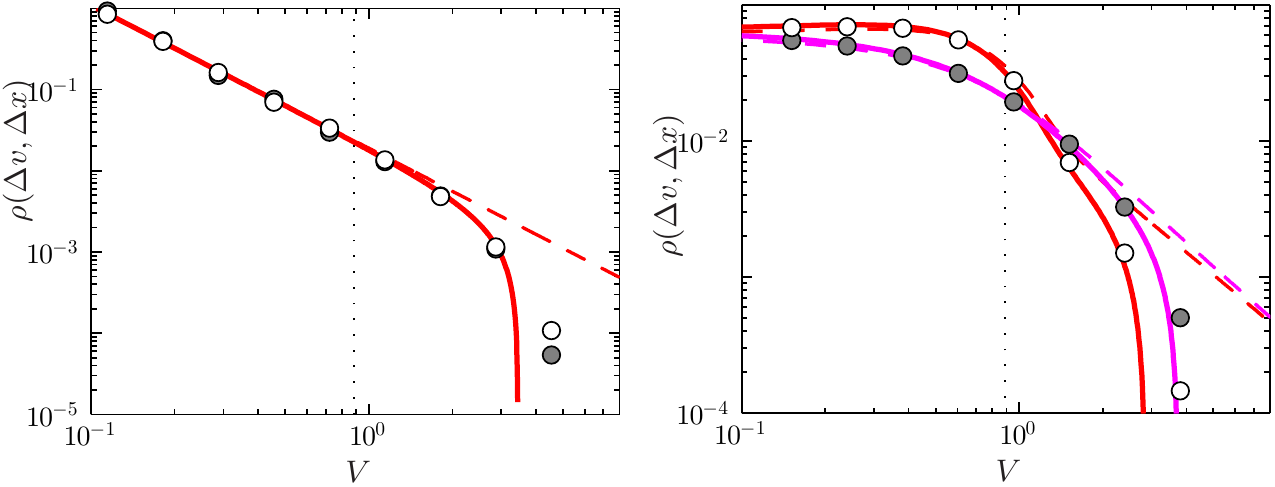}
\caption{\label{fig:distributions1d_boundary}
Zoom in of tails of the distribution (data from left panel in \Figref{distributions1d}).
Left: $\Delta x=10^{-6}$. Right: $\Delta x=0.5$ (not shown in \Figref{distributions1d}).
Solid lines correspond to theory \eqnref{rho_finiteX_ansatz}, dashed lines show the $\R\to 0$-theory, \Eqnref{rho1d_snallR_final},
plotted in \Figref{distributions1d}. Black dotted lines show cut-off $\Delta v_{\rm c}=\xstar\zstar$ with $\xstar=\sqrt{2/3}$ for the model in \Secref{model}.
}
\end{figure}

The final result \eqnref{rho_finiteX_ansatz}
for the distribution of relative velocities at small separations
is shown in Fig.~\ref{fig:distributions1d_boundary}, compared
with results of numerical simulations of Eq.~(\ref{eq:deutsch})
for $\epsilon^2=3$. We see that \eqnref{rho_finiteX_ansatz}
improves upon \eqnref{rho1d_snallR_final}. \Eqnref{rho_finiteX_ansatz}
describes correctly how the power-law tails in \eqnref{rho_finiteX_ansatz}
are cut off.
The differences between the solutions \eqnref{rho_finiteX_ansatz} and \eqnref{rho1d_snallR_final} are most significant for $\R$ and $\V$ close to their cut-off values $\R\sim 1$ or $\V\sim\zstar$.
But it is also clear that \eqnref{rho_finiteX_ansatz}
fails to converge for very large values of $\V$.

\section{Conclusions}
\label{sec:C}

In this paper we have computed the joint distribution of spatial separations
and relative velocities of inertial particles suspended in incompressible,
turbulent and randomly mixing flows, \Eqnref{rho_asym_generaldim}.
This result is based on matching asymptotically known forms
of the distribution, and takes into account fractal clustering as well
as the occurrence of singularities (caustics) at finite Kubo numbers. The distribution is parameterised in terms of two parameters,
the phase-space correlation dimension $D_2$,
and a matching parameter $\zstar$.

Our most important conclusion is that
the form of the distribution of relative velocities
at small separations \eqnref{rho_asym_generaldim}
is expected to be universal, that
is independent of the particular properties of the random
or turbulent flow (it is assumed that the flow is isotropic,
homogeneous and incompressible).
In particular, the universal result also applies to white-noise flows.
This explains why certain aspects of the white-noise results derived in~\cite{Gus11b} are good approximations also when $\ku$ is of order unity, as well
as in turbulent flows.

A universal feature of the distribution of relative velocities $\Delta\ve v$,
\eqnref{rho_asym_generaldim}, is its power-law form
at small separations $\R\ll 1$:
\begin{equation}
\eqnlab{V_Power}
\rho(\Delta\ve v,\R) \sim \V^{D_2-2d}\,,
\end{equation}
valid provided that $\V=|\Delta\ve v|$ is large, but not too large.

By integration of the universal distribution \eqnref{rho_asym_generaldim} we have found that the distribution of radial relative velocities $\Vr$
at small separations $\R$ obeys a power law too,
\eqnref{rhoVr_asym2_generaldim}:
\begin{equation}
\eqnlab{Vr_Power}
\rho(\Vr,\R) \sim |V_R|^{D_2-d-1}\,,
\end{equation}
provided that $D_2 < d+1$ and that $\Vr$ is large (but not too large).
We have shown that the power-law forms of Eqs.~\eqnref{V_Power} and~\Eqnref{Vr_Power} reflect the presence of caustics in the inertial dynamics of the particles. The exponents in \eqnref{V_Power} and \eqnref{Vr_Power} can be independently determined in an experiment or in direct numerical simulations of inertial particles in turbulent flows. Fitting experimental or numerical results to Eqs.~\eqnref{V_Power} and~\eqnref{Vr_Power} thus constitutes a strong test of the underlying theory  \cite{Wil05,Wil06} of caustics in turbulent aerosols.

The power-law forms of the distributions of relative
velocities found in this paper imply that caustics
make a strong contribution to the  moments
of $|V_R|$. We have found that these
moments approximately take the form \eqnref{mp_generalform}
\begin{equation}
m_p(R) \sim b_p R^{p+D_2-1} + c_pR^{d-1}\,.
\eqnlab{mp_generalform_redisplay}
\end{equation}
for small distances $\R$. The same form was obtained in the
white-noise limit in \cite{Gus11b}.
The first term results predominantly from smooth particle-pair diffusion, the second term corresponds to the singular contribution of caustics.
Our findings imply that
caustics make a substantial contribution
to the collision kernel, in keeping
with the results obtained in \cite{Wil06,Gus11b}.

From the form of the moments \eqnref{mp_generalform_redisplay} with $p=0$ it follows that the phase-space correlation dimension $D_2$ is related to the real-space correlation dimension $d_2$ by
\begin{equation}
\label{eq:dd}
d_2 = \min(D_2,d)\,,
\end{equation}
consistent with the white-noise results described in \cite{Gus11b},
and with the results of numerical simulations described
in \cite{Bec08}. Eq.~(\ref{eq:dd}) implies
that the real-space and phase-space correlation dimensions
are equal for not too large Stokes numbers.

The parameter dependence of the coefficients in $b_p$ and $c_p$ in \eqnref{mp_generalform_redisplay} is system dependent.
If non-ergodic effects are small for the distribution of relative velocities it is possible to relate $b_p$ to $b_0$ [see \Eqnref{mp_smooth_smallR}]. For large values of $\st$ it is possible to relate $c_p$ to $c_0$ [see Eqs.~\eqnref{mp_inertial_range} and~\eqnref{mp_largest_scale}].
This allows to calculate the particle-velocity structure functions $S_p\equiv m_p/m_0$, in good agreement
with recent results of direct numerical simulations of particle suspended in turbulent flows \cite{Bec10,Cen11,Sal12}.

In order to obtain an approximate expression for the collision rate,
it is necessary to evaluate the moment $m_1$
[see \Eqnref{collrate_approx}].
In order to find the $\st$-dependence of $m_1$,
the $\st$-dependence of $b_0$ and $c_0$ must be calculated.
For single-scale flows in the limits stated above, $b_0$ is obtained in \Eqnref{b0} and $c_0$ is approximately independent of $\st$. For multi-scale flows, the $\st$-dependence in the coefficients $b_0$ and $c_0$ is not yet known,
neither analytically nor from direct numerical simulations
of particles suspended in turbulent flows.

Last but not least we have shown how the matching parameters
(the correlation dimension $D_2$ and $z^\ast$) can be calculated
from first principles in one spatial dimension, in the white-noise
limit. The results are in good agreement with numerical results for the distribution of relative velocities (\Figref{distributions1d}).
In one spatial dimension we have also shown how the power-law form
of the distribution of relative velocities is cut off in the far tails.
Our results are in good      
agreement with numerical simulations (\Figref{distributions1d_boundary}).

In summary, our analytical results provide a
rather complete description of the distribution
of relative velocities of inertial
particles in random velocity fields, and of the
universal properties of this distribution
for turbulent aerosols. We expect
that our results will make it possible
to obtain an accurate analytical
parameterisation of the collision
kernel of identical particles colliding
in turbulent flows.

\section*{Acknowledgements}
Financial support by Vetenskapsr\aa{}det, by
the G\"oran Gustafsson Foundation for Research in Natural Sciences and Medicine, and by the EU
COST Action MP0806 on \lq Particles in Turbulence' are
gratefully acknowledged. The numerical computations
were performed using resources provided by C3SE and
SNIC.

\appendix
\section[Calculation of ak(z)]{Calculation of $\alpha_k(\zeta)$}
\applab{ak_calculation}
In this appendix we show how to
recursively solve \Eqnref{start_FP} to find $\alpha_k(\zeta)$.

We have $\alpha_0(\zeta)=Z_{D_2}(\zeta)$ for $k=0$ in \Eqnref{Alt_FP_smallR_recursion}.
Consider the diffusion constant \eqnref{sqrtDX_finiteX_expansion}.
If the range where ${\cal D}(R) \sim \epsilon^2 R^2$ extends to infinity (i.e. if $\beta_i=0$ in \eqnref{sqrtDX_finiteX_expansion} for $i>0$),
then $Z_{D_2}(\zeta)$ is the full solution and  $\alpha_k(\zeta)=0$ for $k>0$.
In this case \Eqnref{start_FP} reduces to \Eqnref{Zmu} with $\mu=D_2+2k$ and solution $\alpha_k(\zeta)=Z_{D_2+2k}(\zeta)$.
In \Ssecref{FP1d_etainf} we discussed the symmetry of the problem under exchanging the particles in a pair. It follows from this symmetry that each $\alpha_k(\zeta)$ must be symmetric in $\zeta$ as $\zeta\to\pm\infty$.
But the spectrum of $\mu$ such that $Z_\mu$ is symmetric, plotted in \Figref{muc}, does not allow for solutions $Z_{D_2+2k}(\zeta)$ (unless $\epsilon\to\infty$).
In conclusion, in the limit of $\R\to 0$ we obtain the solution found in \Ssecref{FP1d_etainf}, $\rho(\zeta,\R)=Z_{D_2}(\zeta) R^{D_2-1}$.

The equations for each $\alpha_k(\zeta)$ in \eqnref{Alt_FP_smallR_recursion} consist of one part that depends upon $\alpha_i(\zeta)$ with $i<k$, and one part identical to the equation \eqnref{Zmu} for $Z_{\mu}$, with $\mu=D_2+2k$.
The form of the $Z_\mu$-solutions motivates us to search for solutions that behave as power laws for large values of $|\zeta|$.
An expansion for large values of $|\zeta|$ in \eqnref{Alt_FP_smallR_recursion}
shows that
\begin{align}
\alpha_k(\zeta)\sim\alpha_k^\pm\,\,(\pm\zeta)^{D_2-2+\delta_k}
\quad\mbox{ as $\zeta\to\pm\infty$}\,.
\end{align}
When $k=0$ we have $\delta_0=0$ from \eqnref{asy} and when $k>0$, $\delta_k$ satisfies either $\delta_k=\delta_{k-1}$ or $\delta_k=2k$.
Now consider the ansatz \eqnref{rho_finiteX_ansatz} for large values of $|\zeta|$ and small values of $\R$ (so that $\zeta\approx z=\Delta v/\Delta x$):
\begin{equation}
\rho(\Delta v,\Delta x)\sim\sum_{i=0}^\infty \alpha_i^\pm\,\,(\pm \Delta v)^{D_2-2+\delta_i}\R^{2i-\delta_i}\,.
\eqnlab{sum}
\end{equation}
When $\Delta x=0$, $\rho(\Delta v,0)$ should drop to zero for large values of $\V$. But since $2k\ge \delta_k\ge \delta_{k-1}\ge 0$,
all orders in $\Delta v$ must be included in the sum \eqnref{sum}
to ensure convergence at $\Delta x= 0$.
This implies the condition $\delta_k=2k$, which
ensures that the factor $\R^{2i-\delta_i}$ does not cut off
the sum \eqnref{sum}.
Particle-interchange symmetry requires that the distribution at $\Delta x=0$
\begin{equation}
\rho(\Delta v,\Delta x=0)\sim\sum_{i=0}^\infty \alpha_i^\pm\,\,(\pm \Delta v)^{D_2-2+2i}
\end{equation}
is  symmetric in $\Delta v$. It follows that
$\alpha_k^+=\alpha_k^-$. For $k=0$ this condition
corresponds to \eqnref{A}.

The last step consists of determining the coefficients
$\alpha_k^-$. Once these coefficients are known, the equations
\Eqnref{Alt_FP_smallR_recursion} can be solved in the same way
as \Eqnref{Zmu}.
We need to find a boundary condition to determine the coefficients
$\alpha_k^-$ so that $\alpha(\zeta)$ is symmetric as $\zeta\to\pm\infty$.
When $k=0$, the solution $\alpha_0(\zeta)=Z_{D_2}(\zeta)$ is always symmetric independently of the value of the coefficient $\alpha^-_0$. We set this global normalisation factor to unity, $\alpha_0^-=1$. When $k>0$,
we can find the values of $\alpha_k^-$ so that $\alpha(\zeta)$ is
symmetric by a numerical shooting method.
A second more efficient possibility to find $\alpha_k^-$ is the
following. We write
\begin{align}
\alpha_k(\zeta)=\frac{Z_{D_2+2k}(\zeta)}{A^-_{D_2+2k}}\alpha_k^-+C_k(\zeta)\,.
\eqnlab{alfak_ansatz}
\end{align}
Here $Z_{D_2+2k}(\zeta)$ is the solution \eqnref{Z2} of \Eqnref{Zmu} with $\mu=D_2+2k$ and $z=\zeta$. The function $C_k(\zeta)$ remains to be determined.
Consider large negative values of $\zeta$.
The asymptotic behaviour of $\alpha_k(\zeta)\sim\alpha_k^-(-\zeta)^{D_2-2+2k}$ matches that of $Z_{D_2+2k}(\zeta)\sim A_{D_2+2k}^-(-\zeta)^{D_2-2+2k}$. 
Inserting this law into \eqnref{alfak_ansatz} shows that the left tail of $C_k(\zeta)$ must be of an order smaller than $D_2-2+2k$ (although the right tail of $C_k(\zeta)$ may be of the order $D_2-2+2k$).
Expanding $\alpha_k(\zeta)$ and $Z_{D_2+2k}(\zeta)$ to lower orders for large negative values of $\zeta$ shows that the asymptotic behaviour of $C_k(\zeta)$
is
\begin{equation}
C_k(\zeta)\sim(D_2-3+2k)\alpha_{k-1}^-\beta_1(-\zeta)^{D_2-4+2k}\,.
\end{equation}
Now consider large positive values of $\zeta$. Because of the particle-interchange symmetry we must require:
\begin{equation}
\left .
\begin{array}{rl}
\alpha_k(\zeta)&\sim\alpha_k^-\zeta^{D_2-2+2k}\,,\nn\\
Z_{D_2+2k}(\zeta)&\sim A^+_{D_2+2k}\zeta^{D_2-2+2k}\nn\\
C_k(\zeta)&\sim\gamma_k\zeta^{D_2-2+2k}\,.\nn
\end{array}
\right \} \quad \mbox{as $\zeta \rightarrow \infty$\,.}
\label{eq:asympto}
\end{equation}
Here $A^+_{D_2+2k}$ is determined from \eqnref{Z2}.
The coefficient $\gamma_k$ is determined as follows.
We insert $\alpha_k(\zeta)$,  \Eqnref{alfak_ansatz},  into
\Eqnref{Alt_FP_smallR_recursion} and solve the resulting
equation. 

\begin{figure}
\includegraphics[width=13.5cm,clip]{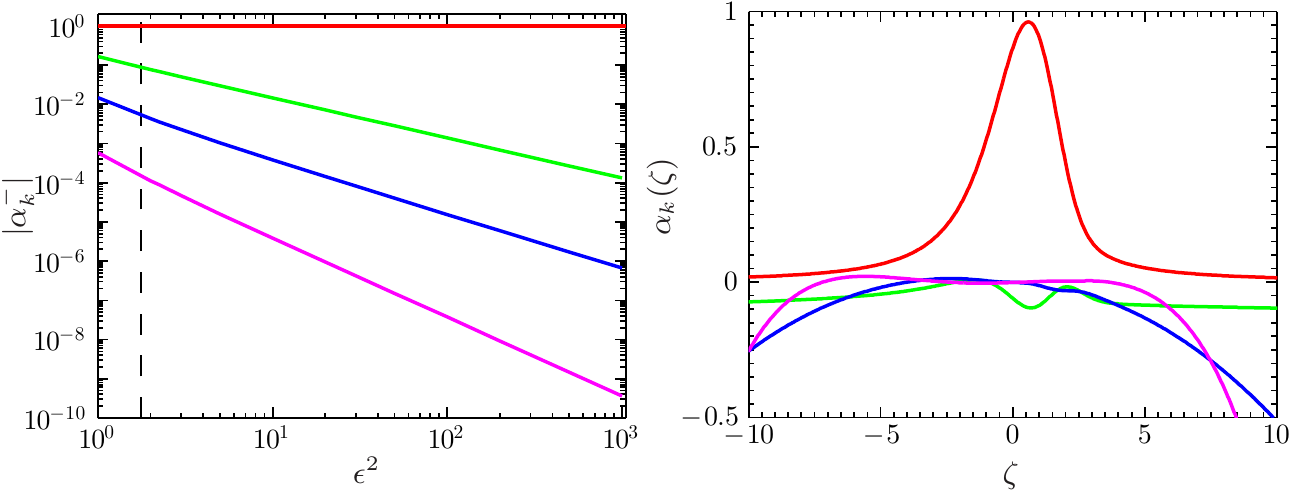}
\caption{\label{fig:alphak}
Left: Coefficients $\alpha_k^-$ obtained by evaluation of \Eqnref{acoeffs_boundary} as a function of $\epsilon^2$,
for $k=0$ (red), $k=1$ (green), $k=2$ (blue) and $k=3$ (magenta).
Right: Solutions $\alpha_k(\zeta)$ to \Eqnref{Alt_FP_smallR_recursion} as a function of $\zeta$ for $\epsilon^2=3$.
}
\end{figure}

Substituting the large-$\zeta$ asymptotes (\ref{eq:asympto})
into \Eqnref{alfak_ansatz} we solve for $\alpha_k^-$ to find
\begin{align}
\alpha^-_k=\frac{\gamma_k}{1-A^+_{D_2+2k}/A^-_{D_2+2k}}
\eqnlab{acoeffs_boundary}
\end{align}
for $k>0$.
This concludes our calculation of the functions $\alpha_k(\zeta)$.
The result is shown (for $k=0,\ldots,3$) in
\Figref{alphak} (right panel). The corresponding
coefficients $\alpha_k^-$ are shown
in the left panel of \Figref{alphak} as a function of $\epsilon^2$.

Substituting $\alpha_k(\zeta)$ into
\Eqnref{rho_finiteX_ansatz} yields the desired approximation
of the tails of the joint distribution of
separations and relative velocities in one spatial dimension,
in the white-noise limit. The result is shown in \Figref{distributions1d_boundary}
and discussed in the main text.

% \bibliographystyle{tJOT}
% \bibliography{refs}

\end{document}